\documentclass[
	aps, prd, reprint,
	10pt, notitlepage, a4paper,
        floats, floatfix,
	amsmath, amssymb, amsfonts, eqsecnum,
	superscriptaddress,
	showpacs, showkeys,
	nofootinbib,
 	longbibliography,
]{revtex4-2}

\usepackage[usenames,dvipsnames]{xcolor}
\usepackage{xspace} 
\usepackage{bm} 
\usepackage[utf8]{inputenc}
\usepackage{mathrsfs}
\usepackage{graphicx}
\usepackage[normalem]{ulem}
\graphicspath{ {Figures/} }

\xdefinecolor{mylinkcolor}{rgb}{0,0,0.5}
\usepackage[
	bookmarksnumbered, bookmarksopen, bookmarksopenlevel=2,
	breaklinks=true,
	colorlinks=true, filecolor=mylinkcolor, citecolor=mylinkcolor,
	linkcolor=mylinkcolor, urlcolor=mylinkcolor, menucolor=mylinkcolor,
]{hyperref}

\def\ee{\end{equation}}
\def\eea{\end{eqnarray}}
\def\be{\begin{equation}}
\def\bea{\begin{eqnarray}}
\def\p{\partial}
\def\lhat{\hat{\ell}}
\def\dhat{\hat{d}}
\allowdisplaybreaks

\newcommand{\GCK}[1]{\textcolor{blue}{#1}}

\begin{document}

\newcommand{\AEI}{\affiliation{Max-Planck-Institute for Gravitational Physics (Albert-Einstein-Institute),\\ Am M{\"u}hlenberg 1, 14476 Potsdam-Golm, Germany, European Union}}
\newcommand{\UU}{\affiliation{Institute for Theoretical Physics, Utrecht University,\\ Princetonplein 5, 3584 CC Utrecht, Netherlands, European Union}}

\title{Tidal response from scattering and the role of analytic continuation}
\date{\today}

\author{Gast\'on Creci}\email{g.f.crecikeinbaum@uu.nl}
\UU
\author{Tanja Hinderer}
\UU
\author{Jan Steinhoff}\AEI

\begin{abstract}
The tidal response of a compact object is a key gravitational-wave observable encoding information about its interior. This link is subtle due to the nonlinearities of general relativity. We show that considering a scattering process bypasses challenges with potential ambiguities, as the tidal response is determined by the asymptotic in- and outgoing waves at null infinity. As an application of the general method, we analyze scalar waves scattering off a nonspinning black hole and demonstrate that the low-frequency expansion of the tidal response reproduces known results for the Love number and absorption. In addition, we discuss the definition of the response based on gauge-invariant observables obtained from an effective action description, and clarify the role of analytic continuation for robustly (i) extracting the response and the physical information it contains, and (ii) distinguishing high-order post-Newtonian corrections from finite-size effects in a binary system. 
Our work is important for interpreting upcoming gravitational-wave data for subatomic physics of ultradense matter in neutron stars, probing black holes and gravity, and looking for beyond-standard-model fields. 
\end{abstract}

\maketitle

\makeatletter
\def\l@subsubsection#1#2{}
\makeatother

\tableofcontents

\section{Introduction}

The ever-increasing number of gravitational-wave detections of merging binary systems ~\cite{LIGOScientific:2018mvr,LIGOScientific:2020ibl,LIGOScientific:2021usb,LIGOScientific:2021qlt} has revealed a wealth of new insights and provided an unprecedented tool for fundamental physics, astrophysics, and cosmology~\cite{LIGOScientific:2020tif,LIGOScientific:2020kqk,LIGOScientific:2018cki,LIGOScientific:2017adf}. In particular, gravitational waves encode unique information about the nature and interiors of compact objects. During the clean, cumulative binary inspiral epoch, these imprints arise from spin-induced deformations and a variety of tidal interactions. Tidal effects are especially interesting because they correspond to the excitation of isolated quasinormal modes of the compact objects driven by the companion's time-varying tidal field due to the orbital motion. The dominant tidal signatures in gravitational waves depend on the objects' internal structure through a characteristic tidal deformability parameter, first measured for the binary neutron star event GW170817~\cite{LIGOScientific:2018cki}. Determining tidal parameters such as deformability is of major interest for understanding the long-sought properties of matter at supranuclear densities in neutron stars~\cite{osti_1296778, NuPECC}, probing the nature of black holes and constraining quantum corrections to their horizons~\cite{Barack:2018yly}, tests of gravity~\cite{Yagi:2013awa,Doneva:2017jop}, and looking for beyond-standard-model fields in the cosmos~\cite{Cardoso:2019rvt,DeLuca:2021ite}. 

In the next years, the gravitational-wave detectors will continue to increase in sensitivity~\cite{KAGRA:2020npa}, discovering more diverse populations of compact objects and performing higher accuracy studies of nearby events. Envisioned next-generation detectors~\cite{Bailes:2021tot,Maggiore:2019uih,Reitze:2019iox} will improve upon the sensitivity of current instruments by an order of magnitude and enable precision physics with gravitational waves. To realize this science potential requires accurate theoretical models, which play a crucial role since the data analysis cross-correlates templates with the data~\cite{LIGOScientific:2019hgc,Cutler:1994ys}. For the events analyzed to date, systematic uncertainties in the modeling and interpretation of the signals have been subdominant compared to the statistical errors, as far as could be quantified~\cite{LIGOScientific:2020ibl}. As the statistical errors decrease in the near future, however, shortcomings in the modeling will become more prominent. There is thus an urgent need to address lingering theoretical challenges that impact our interpretation of the gravitational wave signals, especially with regard to finite-size effects. 

The relativistic tidal deformability characterizing the dominant finite-size effects in the gravitational waves is defined as the ratio of the induced multipolar deformation of a compact object's exterior spacetime to the perturbing tidal field of the companion~\cite{Flanagan:2007ix}. In the static limit, when the perturbing frequency is far below a mode resonance~\cite{Chakrabarti:2013lua}, it reduces to the relativistic tidal Love numbers~\cite{1909RSPSA..82...73L,Flanagan:2007ix}. Its computation is based on perturbation theory for compact objects~\cite{Damour:2009vw,Binnington:2009bb, Chakrabarti:2013lua,Hinderer:2007mb}, with many different examples of compact objects studied to date, e.g.~\cite{Damour:2009vw,DeLuca:2021ite,Cardoso:2017cfl,Pereira:2020jgv,Hinderer:2009ca,Ciancarella:2020msu,Cardoso:2019rvt,Han:2018mtj,Postnikov:2010yn,Sennett:2017etc}.
Recent work~\cite{Gralla:2017djj,Chia:2020yla, Kol:2011vg,Hui:2020xxx, LeTiec:2020bos,LeTiec:2020spy,Poisson:2020vap,Kim:2020dif,Hui:2021vcv,Charalambous:2021kcz,Poisson:2021yau} has revealed a number of intricacies in addressing questions such as: How are the measurable tidal signatures in gravitational waves related to the response of a compact object obtained from perturbation theory? Is the tidal deformability identically zero for all black holes in general relativity and four spacetime dimensions?

Despite recent progress, several concerns remain, for instance, about degeneracies of tidal effects with high post-Newtonian (PN) order terms describing relativistic corrections to the dynamics of point masses. The problem is the following, as described in detail in~\cite{Gralla:2017djj}. The tidally induced multipole moments can be defined from the exterior spacetime of the perturbed compact object in asymptotically cartesian mass-centered coordinates~\cite{Thorne:1980ru}, which is equivalent to other definitions of spacetime multipole moments~\cite{1983GReGr..15..737G}. For instance, the metric component $g_{tt}$ (the analog of the Newtonian gravitational potential) has an asymptotic expansion at large distances $r\to\infty$ from a nonspinning compact object of the form
$(1-g_{tt})=-2M/r-3Q/ r^3+\ldots +{\cal E} \, r^2+\ldots $,
where the omissions denote higher orders in $1/r$ and $r$.
The quantity $Q$ is the quadrupole moment contracted with two copies of a unit vector, and ${\cal E}$ is the similarly contracted quadrupolar tidal field. The multipole moments of the object are associated with the asymptotically decaying series in $1/r$, while the external field corresponds to the growing terms in $r$. However, ambiguities arise when the two series overlap. For instance, this occurs when the tidal field is only known to some order in its PN expansion having the schematic form ${\cal E}={\cal E}^{\rm N}[1+\ldots+\delta_5/r^5+O(r^{-6})]$, where ${\cal E}^{\rm N}$ is the Newtonian tidal field, and schematically, an $n$th order PN correction contributes a power of $1/r^n$. For a binary system, the coefficient $\delta_5$ remains unknown at present. A fractional change in ${\cal E}$ at $O(r^{-5})$  introduces a net $1/r^3$ contribution to the expansion of $g_{tt}$, thus changing the multipole moment ${\cal Q}$ by an amount proportional to $\delta_5$. This degeneracy between PN and multipole effects is discussed in more detail in~\cite{Kol:2011vg,Damour:2009vw,Pani:2015hfa,Damour:2009va,Chakrabarti:2013lua, Gralla:2017djj,Flanagan:1997fn}.

In this paper, we clarify that an unambiguous distinction between finite-size effects and high order PN corrections is achieved by using analytic continuation. We show that working in generic spacetime dimensions and/or multipole moments avoids degeneracies and manifestly separates these two kinds of physical effects. In addition, we emphasize that potential ambiguities arising from inspecting the metric in specific coordinates, as explained above, are avoided by defining the response based on gauge-invariant quantities such as the binding energy as a function of frequency or the waveform~\cite{Gralla:2017djj}. In related contexts, though at fixed spacetime dimension and multipole orders, the binding energy has long proved useful for identifying multipole moments~\cite{Ryan:1995wh,Pani:2015hfa}.

The binding energy can be derived from an effective action, which is a highly useful tool in physics~\cite{Georgi:1993mps,Goldberger:2007hy}, and in particular for calculating finite-size effects at the orbital  scale~\cite{Goldberger:2004jt,Bini:2012gu,Steinhoff:2016rfi,Kol:2011vg, Goldberger:2007hy,Goldberger:2005cd,Chakrabarti:2013xza}; for broader review articles see e.g.~\cite{Levi:2018nxp,Porto:2016pyg}. The effective action describes the compact objects by skeletonized center-of-mass worldlines~\cite{Dixon:1970zza} augmented with multipole moments. Couplings of the multipole moments to the ambient spacetime curvature and possible internal dynamics encapsulate the finite size effects.
An important feature of this approach is that the information about the object encoded, for instance, in the values of the coupling coefficients must be matched to a detailed microphysical description, often based on calculations in perturbation theory. In this paper, we demonstrate the advantages of establishing this connection by recasting the problem into a scattering calculation and matching the frequency-dependent response, instead of working with stationary perturbations and specializing to the static limit as in many previous studies~\cite{Kol:2011vg,Hui:2020xxx, Chia:2020yla,LeTiec:2020bos,LeTiec:2020spy,Damour:2009vw,Binnington:2009bb,Hinderer:2007mb}. The matching of scattering states enables us to identify the black hole's tidal response from in- and outgoing waves at null infinity using double-null Bondi coordinates~\cite{Bondi:1962px}. These coordinates are defined from light-cone congruences, whose intrinsic geometric meaning is further reviewed in~\cite{Blanchet:2020ngx}.

As a first step toward the more complicated gravitational case we consider here a scalar field model, which nevertheless captures a number of important features of potential ambiguities. This  scenario has also been studied in~\cite{Kol:2011vg,Hui:2020xxx}, with the differences being that we consider wave scattering and keep the frequency-dependence in the response. 
For black holes, this response also includes dissipative effects due to absorption, which can also be described in effective field theory~\cite{Goldberger:2005cd,Porto:2007qi,Endlich:2015mke,Endlich:2016jgc,Goldberger:2012kf,Goldberger:2020wbx,Goldberger:2020fot,Goldberger:2019sya,Goldberger:2020geb}. We verify that the low-frequency expansion of the black hole's frequency-dependent scalar tidal response obtained from scattering recovers the expected results of a vanishing tidal deformability~\cite{Kol:2011vg} and the known absorption cross section~\cite{Page:1976df} at zeroth and linear order in the frequency respectively.

The frequency-dependent response based on scattering was also analyzed in the context of neutron stars in~\cite{Chakrabarti:2013lua}. The new aspects of this paper are that we substantially expound on the methodology, introduce more rigorous identifications at null infinity between the microphysical results and the effective action description, and elaborate on a number of insights. Complementary aspects of wave scattering and absorption by compact objects have been studied in e.g.~\cite{Bernuzzi:2008rq,Stratton:2019deq,Chen:2011jgd,Bautista:2021wfy,Stratton:2020cps,Stratton:2019deq,Leite:2017zyb,Dolan:2017rtj,Dolan:2017zgu,Pons:2001xs}. 

Recent efforts have also highlighted the convenience of using analytic continuation to complex angular momentum for efficiently extracting physical properties of perturbed black hole spacetimes~\cite{OuldElHadj:2019kji,Folacci:2018sef,Folacci:2019vtt,Folacci:2019cmc}.
In this paper, we further examine the role of analytic continuation both in the spacetime dimension and the angular momentum number for the different stages in the calculation of the tidal response. Notably, we demonstrate that in contrast to the crucial role of analytic continuation for obtaining the response in stationary scenarios~\cite{Chia:2020yla, Kol:2011vg,Hui:2020xxx, LeTiec:2020bos,LeTiec:2020spy}, using scattering states might eliminate the need for analytic continuation to match the coefficients in the effective action and the need for high-PN order calculation of observables.  This is advantageous, for instance, for future numerical calculations of the microscopic response that capture the full frequency dependence and are applicable to any compact object. 

This paper is organized as follows. We discuss the gauge-invariant binding energy as a function of frequency of a binary system in Sec.~\ref{sec:bindingE}, highlighting the crucial role of analytic continuation in the spacetime dimension and multipole order to simplify the separation between finite-size effects and PN corrections to the point-mass dynamics. In Sec.~\ref{sec:scattering} we calculate the tidal response function based on the asymptotic scalar wave scattering states extracted at null infinity. We first set up an effective action description, which defines the tidal response that is imprinted in observables, and derive its relation to the in- and outgoing complex wave amplitudes. Next, we connect these wave amplitudes with the detailed properties of a perturbed nonspinning black hole computed from perturbation theory. We elucidate the matching procedures, the role of analytic continuation in the process, and discuss the advantages of scattering over considering stationary scenarios for bypassing several subtleties and giving access to dynamical tides. Section~\ref{sec:discussion} summarizes our results and insights gained from the calculations, and Sec.~\ref{sec:conclusion} contains the conclusions. The Appendices contain additional technical details and mathematical identities used in this work. 

The notation and conventions are the following. Greek letters $\alpha,\beta,\dots$ denote spacetime quantities, Latin indices $i,j,\dots$ denote spatial components. The notation for the covariant derivative is $\nabla_\mu$ and for the partial derivative it is $\p_\mu$. Capital-letter superscripts denote a string of indices on a symmetric and trace-free (STF) tensor, e.g for a unit vector $n^{L=2}=n^{ij}=n^in^j-\frac{1}{3}\delta^{ij}$, see ~\cite{Thorne:1980ru}. We use the Einstein summation convention on all types of indices, i.e. repeated indices are summed over. We work in generic spacetime dimensions $d$, and use the shorthand $\dhat=D-2=d-3$, where $D$ is the number of \emph{spatial} dimensions, such that four-dimensional spacetime corresponds to $\dhat=1$. We also define $\lhat=\ell/\dhat$, with $\ell$ the multipolar order. Throughout the paper, we work in units where the speed of light is unity but we explicitly keep the gravitational constant $G_N$.

\section{Gauge-invariant binding energy of a binary system}
\label{sec:bindingE}

In this section, we discuss how potential ambiguities -- or rather, technical difficulties -- in the identification of tidal deformabilities can arise~\cite{Damour:2009va,Damour:2009vw,Binnington:2009bb,Kol:2011vg,Gralla:2017djj} and be overcome by using gauge-invariant quantities~\cite{Gralla:2017djj} and analytic continuation e.g.~in dimension or multipolar order~\cite{Kol:2011vg,Chakrabarti:2013lua,Steinhoff:2014kwa,LeTiec:2020spy,LeTiec:2020bos,Hui:2020xxx,Chia:2020yla}. Specifically, we demonstrate that the circular-orbit binding energy connects the tidal deformabilities to an observable, while analytic continuation enables discriminating tidal from nontidal PN contributions.
For clarity and conciseness of the expressions, we focus on the static tides and nonspinning objects.

We first consider tidal effects in Newtonian gravity to derive an explicit effective action for the orbital dynamics in arbitrary spacetime dimensions. As in four dimensions, the action derived in this way depends on the compact object's microphysics only through the tidal deformability. It also describes fully relativistic compact objects at large separation, provided that one interprets the tidal deformability as the relativistic parameter~\cite{Chakrabarti:2013xza,Flanagan:2007ix, Steinhoff:2016rfi}.
We then discuss tidal effects in the binding energy in relation to nontidal PN contributions.

\subsection{Expansions of the potential: Multipole and tidal moments}
We consider an extended object (labeled $A$) and a point-mass companion (labeled $B$). The gravitational potential $U$
is a solution to the $D$-(spatial)dimensional Poisson equation:
 \be\label{eq: DPoissonEq}
 \nabla^{2(D)}U=-\Omega_DG_N^{D}\rho_D~,
\ee
where $\Omega_D=2\pi^{D/2}/\Gamma(D/2)$ is the volume of the $(D-1)$-hypersphere, $G_N^{D}$ and $\rho_D$ are the $D$-dimensional gravitational constant and mass density respectively, and $\nabla^{2(D)}$ is the $D$-dimensional Laplacian. In $D$-dimensions, the gravitational constant $G_N^{(D)}$ has dimensions $[{\rm s}]^{-2}[{\rm kg}]^{-1}[{\rm m}]^{D}$. 
From now on, we will omit the label $D$ and use the notation $\dhat=D-2$, such that $\dhat=1$ corresponds to $3$ spatial dimensions. In the exterior of a single body, the solution of \eqref{eq: DPoissonEq} reads
\begin{align}
\label{eq:Uselfexp}
U_A&=\frac{G_N M_A}{\dhat|{x^i}-{z^i}_A|^{\dhat}}\nonumber\\
&+\sum_{\mathcal{\ell}=2}^{\infty} \frac{1}{\ell!}\frac{(2\ell+\dhat-2)!!}{\dhat!!}\frac{G_NQ_Ln^{L}}{|{x^i}-{z^i}_A|^{\ell+d}},
\end{align}
with $z^i_{A}$ the center of mass position of the body $A$, and we have separated out the point-mass result in the first term. The moments $Q^L$ are the $D$-dimensional Newtonian
source multipole moments defined as integrals over the mass density
\begin{align}
Q^L=\int{d^{\dhat+2}x' \rho_A(t,{{x'}^i})(x'-z_A)^L}.\label{eq:QNewt}
\end{align}
In a binary system, the total potential near the extended body $A$ also has a contribution from the fact that the potential due to the companion varies over $A$'s mass distribution. The potential felt by $A$ due to the companion can be expressed as
\begin{align}
\label{eq:Uextexpand}
U_A^{\rm ext}=-\sum_{\ell=0}^{\infty}\frac{1}{\ell!}(x-z_A)^L\mathcal{E}_L~,
\end{align}
where ``ext" indicates that the source is external to the object. We have also introduced the tidal tensor
\begin{align}
\mathcal{E}_L=G_N(-1)^{\ell+1}\frac{(2\ell+\dhat-2)!!}{\dhat!!}\frac{M_B}{r_{AB}^{\ell+\dhat}}n_{AB}^{L},
\end{align}
where $M_B$ is the mass of the companion, $r_{AB}$ denotes the distance between the bodies, and $n_{AB}^i=(x_A^i-x_B^i)/r_{AB}$ is a unit vector.

\subsection{Effective action for extended objects in Newtonian gravity}
\subsubsection{Lagrangian for the binary dynamics}
Next, we compute the Lagrangian, $\mathcal{L}=T-V$, with $T=T_A+T_B$ and $V=V_A+V_B$ the kinetic and potential energy obtained by generalizing the results of~\cite{Vines:2010ca} to arbitrary dimensions
\begin{subequations}
\label{eq:TandV}
\begin{align}
&T_A=\frac{1}{2}\int_Ad^{\dhat+2}x\,\rho_A\,\dot{z}_A^2+T^{\rm int}_A,\\
&V_A=-\frac{1}{2}\int_Ad^{\dhat+2}x\, \rho_A\, U_A^{\rm ext}+V_A^{\rm int}.\label{eq:VAdef}
\end{align}
\end{subequations}
Here, $T^{\rm int}_A$ and $V^{\rm int}_A$ are the internal kinetic and potential energy, which we will specify below. The contributions from object $B$ are similar to~\eqref{eq:TandV} but only the point-mass terms are nonvanishing. Substituting the expansion of the potential~\eqref{eq:Uextexpand} into~\eqref{eq:VAdef} and using the definition~\eqref{eq:QNewt} leads to the
 $\dhat$-dimensional Newtonian action with tidal effects 
\be
\label{eq:actiongrav1}
S_{\rm Newt}=S_{\rm pm}^{\rm Newt}+\int{dt}\left[-\sum_{\ell=2}^{\infty}\frac{1}{\ell!}Q_L\mathcal{E}^{L}+\mathcal{L}^{\rm int}_A\right],
\ee
with $S_{\rm pm}^{\rm Newt}$ the Newtonian action for point mass dynamics
\be
S_{\rm pm}^{\rm Newt}=\int{dt}\left[\frac{1}{2}\mu{v^2}+\frac{1}{\dhat}\frac{G_N\mu{M}}{r^{\dhat}}\right].
\ee
Here, $\mu=M_A M_B/M$ and $M=M_A+M_B$ are the reduced and total mass of the binary system, and $\mathcal{L}^{\rm int}_A$ is the Lagrangian for the internal dynamics of the extended body.

\subsubsection{Specializing to the dominant tidal effects}

In Newtonian gravity, the internal dynamics of the multipole moments are directly related to the density perturbations of the matter, and hence the normal modes of oscillation of the object. The fundamental ($f$)-modes have the strongest tidal couplings, and are described by the internal Lagrangian\footnote{This is analogous to a simple harmonic oscillator with $\mathcal{L}=T-V=\frac{1}{2}m\dot{x}^2-\frac{1}{2}kx^2=\frac{1}{2\lambda\omega^2}(\dot{x}^2-\omega^2x^2)$ with $k/m=\omega^2$ and $k=1/(\ell!\lambda_{\ell})$, where the factor $\ell!$ comes from the definition of the tidal deformability. Intuitively, the mass density perturbations due to a companion induce the multipole moments of a initially spherical star. If these perturbations oscillate due to normal modes, the only way they can enter the system is by making the multipoles oscillate. This is the reason why, in this example, $x^2\rightarrow\sum_\ell{Q_LQ^L}$.
}
\be
\label{eq:Lint}
{\cal L}^{\rm int}_A=\frac{1}{2\ell!\lambda_\ell\omega_{0\ell}^2}\left[\dot{Q}_L\dot{Q}^L-\omega_{0\ell}^2Q_LQ^L\right]~.
\ee
The quantities $\omega_{0\ell}$ are the $f$-mode frequencies, with the subscript $0$ indicating that the mode function has no radial nodes, and $\lambda_\ell$ are the tidal deformability coefficients. In the adiabatic limit that the $f$-mode frequency $\omega_{0\ell}$ is much higher than the tidal forcing frequency, which varies on the orbital timescale, the internal Lagrangian reduces to
\begin{align}
\mathcal{L}^{\rm adiab}_{\rm int}=-\frac{1}{2\ell!\lambda_\ell}Q_LQ^L~.
\end{align}
The induced multipole moments $Q^L$ are then related to the tidal moments ${\cal E}_L$ by the equations of motion
\begin{align}
\label{eq:QLlambda}
Q_L^{\rm adiab}=-\lambda_{\ell}\mathcal{E}_L.
\end{align}
It is important to note that~\eqref{eq:QLlambda} is also valid for compact objects in general relativity, when using relativistic definitions of the multipole and tidal moments determined from the exterior spacetime. 

\subsubsection{Reduced effective action}
Using~\eqref{eq:QLlambda} to integrate out the multipole degrees of freedom from the action leads to a reduced action involving only the orbital quantities as dynamical variables
\begin{align}
S^{\rm adiab}&=S_{\rm pm}^{\rm Newt}+\int dt \, \sum_{\ell=2}^{\infty}\frac{\lambda_{\ell}}{2\ell!} {\cal E}_L {\cal E}^L \\
&=S_{\rm pm}^{\rm Newt}+\int dt \, \sum_{\ell=2}^{\infty}\frac{1}{2\ell!}\lambda_{\ell}\Pi_\ell\frac{G_N^2M_B^2}{r^{2\ell+2\dhat}}.
\label{eq:actionbinding2}
\end{align}
To obtain~\eqref{eq:actionbinding2} we used the identity (see Appendix~\ref{app:ContractionNL})
\begin{align}\label{eq:ContrationNL}
n^{L}n_{L}=\frac{(\ell+\dhat-1)!}{(2\ell+\dhat-2)!!(\dhat-1)!!}~,
\end{align}
and defined
\be
\Pi_\ell=\frac{(2\ell+\dhat-2)!!(\ell+\dhat-1)!}{\dhat!!\dhat!}~.
\ee
The information about the internal structure of the compact object is encoded in the action~\eqref{eq:actionbinding2} through the coefficients $\lambda_\ell$. 

\subsection{Tidal effects in the binding energy for circular orbits}

\subsubsection{Computation of the Newtonian binding energy}
As the orbital separation is a coordinate-dependent notion, it is  advantageous to express results in terms of an observable frequency instead. We achieve this by 
expressing the velocity as $v^2=\dot{\phi}^2r^2+\dot{r}^2$ defining the orbital frequency $\Omega$, and considering the equations of motion obtained from~\eqref{eq:actionbinding2} for stable circular orbits $\ddot{r}=\dot{r}=\ddot{\phi}=0$. We solve these perturbatively for the radius as a function of the orbital frequency in the form 
\be
\label{eq:rofomega}
r(\Omega)=\frac{\sqrt{x}}{\Omega}(1+\delta{r}),
\ee
with $\delta r$ denoting the tidal corrections, where 
\be
\label{eq:xdef}
x=(G_N\, M\, \Omega^{\dhat})^{\frac{2}{(\dhat+2)}}.
\ee
We find that the tidal correction in~\eqref{eq:rofomega} is given by
\begin{align}
\delta{r}=\sum_{\ell=2}^{\infty}\lambda_{\ell}\frac{M_B^2\Pi_\ell(\ell+\dhat)}{(\dhat+2)\ell!\mu}\frac{G_N^{\frac{2}{\dhat+2}(\dhat+1-\frac{2\ell}{\dhat})}x^{1+\frac{2\ell}{\dhat}}}{M^{2(\dhat+\ell)/\dhat}}. 
\end{align}
We compute the binding energy as a function of the frequency by reversing the sign of the potential in the Lagrangian~\eqref{eq:actionbinding2}, specializing to circular orbits, and using~\eqref{eq:rofomega} to eliminate $r$ in favor of $\Omega$, or rather $x(\Omega)$. This leads us to
\begin{align}
\label{eq:Eofomega}
&\frac{E(\Omega)}{\mu}=-\frac{1}{2}x\Bigg[\frac{2-\dhat}{\dhat}\nonumber\\&-\sum_{\ell=2}^{\infty}\frac{\dhat(4\hat{\ell}+3)-2}{(\dhat+2)(\hat{\ell}\dhat)!}\Pi_{\ell}\lambda_{\ell}\frac{M_B^2}{\mu M^{2(\hat{\ell}+1)}}x^{2\hat{\ell}+1}G_N^{-2\hat{\ell}}\Bigg] ,
\end{align}
with
\be
\label{eq:lhatdef}\lhat=\ell/\dhat.
\ee

\subsubsection{Discriminating tidal from post-Newtonian (PN) effects}

Having calculated the leading-order tidal effects in the binding energy~\eqref{eq:Eofomega}, we now consider how PN corrections will enter into this expression. In four spacetime dimensions, low PN order fractional corrections to the binding energy scale with integer powers of $x=(G_N\,M\, \Omega)^{2/3}$ in units with the speed of light $c=1$. Terms of $O(x^n)$ correspond to the $n$-PN order. This continues to hold in arbitrary dimensions, as inferred from the explicit calculations of the 1PN Lagrangian in~\cite{Cardoso:2008gn} when using the generalized frequency-variable $x$ that depends on $\dhat$ from~\eqref{eq:xdef}. 
In four dimensions, starting at the fourth PN order, additional terms in the binding energy of the form $x^4\ln(x)$ first appear. These are due to tail effects associated with the scattering of gravitational waves off the spacetime curvature produced by the total mass of the binary and interacting with the system at a later time. Such tail effects are related to the difference in the light cones between Schwarzschild and Minkowski spacetime, as we will discuss in a different context in Sec.~\ref{sec:Schwarzschild}. Because wave propagation is very different depending on the number of spacetime dimensions, we expect that the appearance of the $x^n\ln(x)$ corrections is not a generic feature of PN terms. The effect of such high-order tail terms has not yet been calculated in arbitrary dimensions, thus the more general dependencies remain unknown. Irrespective of the exact scalings of higher PN order effects, it is clear that they are independent of $\lhat$, in contrast to tidal terms. 


Based on the above considerations, the binding energy with tidal and instantaneous PN corrections has the schematic form
\be
\label{eq:Eofomegaschematic}
\frac{E(\Omega)}{E_0}\sim 1+\sum_{\ell=2}^{\infty}\, x^{1+2\hat{\ell}}\,\lambda_{\ell}\, c^{\rm tidal}_{\ell \dhat} +\sum_{\substack{n=2,\\
n\in \mathbb{Z}^+}}^\infty \, x^{n}\, c_{n\dhat}^{\rm PN} +\ldots ,
\ee
where $E_0$ is the Newtonian point-mass result from the first term in~\eqref{eq:Eofomega}, and $c^{\rm tidal}$ and $c^{\rm PN}$ are coefficients that depend on the masses, whose explicit form is not needed here. 


In the standard calculations, where~\eqref{eq:Eofomegaschematic} is specialized to four spacetime dimension $\dhat=1$ and positive integer $\ell$, the tidal terms contribute at orders $x^5, \, x^7, \ldots$, while the PN terms scale as $x, \, \ldots x^5, \ldots$. Consequently, at $O(x^5)$ and higher, there are contributions from both tidal and post-Newtonian terms, where the PN coefficients $c^{\rm PN}_5$ and higher are currently not yet known and are challenging to compute (see, e.g.,~\cite{Bini:2019nra,Blumlein:2020pyo,Foffa:2020nqe}). However, when using analytic continuation keeping $\lhat$ arbitrary, we see from~\eqref{eq:Eofomegaschematic} that tidal terms scale as $x^{(1+2\ell/\dhat)}$, which is manifestly distinct from the instantaneous PN terms that scale as $x^{n}$ with positive integers $n$, irrespective of the dimension or multipolar number.

We note that instead of using analytic continuation in $\lhat$, only one of the analytic continuations in $\dhat$ or in $\ell$ is needed to make the distinction between finite-size and relativistic point-mass effects. For instance, fixing $\dhat=1$ and using analytic continuation in $\ell$ is already sufficient to distinguish tidal terms scaling as $x^{2\ell+1}$ from post-5-Newtonian terms that are expected to scale as $x^5$ and $x^5\ln(x)$. Likewise, specializing to quadrupole tides $\ell=2$ and keeping $\dhat$ arbitrary leads to tidal terms that scale as $x^{1+4/\dhat}$, which is distinct from PN scalings that involve $x^n$ with integer $n$.


The discussion of the tidal signatures in the binding energy also carries over to the tidal imprints on gravitational wave signals. The imprints in the gravitational waves can be estimated by using energy balance as reviewed in~\cite{Blanchet:2013haa}. The energy flux in gravitational waves is computed in a first approximation from the quadrupole formula $\dot E_{\rm GW}\sim \dddot Q^T_{ij}\dddot{Q}^T_{ij} $ (see~\cite{Blanchet:2013haa} for higher order results), which depends on the total quadrupole $Q_{ij}^T$ of the binary system given by the sum of the orbital quadrupole and the tidally induced quadrupoles. Thus, the energy flux also depends on $Q_L$ and hence the tidal response, except with different mass-dependent coefficients than in the binding energy. The combination of these two dependencies on the response ultimately leads to the tidal signatures in the gravitational waves. 

\subsection{Matching the tidal deformability}

 The above results allow us to explore different avenues for extracting the tidal deformabilities from the binding energy of a binary. First, since the circular-orbit binding energy is a gauge- or coordinate-independent observable, one can match our EFT result~\eqref{eq:Eofomegaschematic} to any other result for this quantity obtained in approaches that comprise the full linear tidal response, and solve for the $\lambda_{\ell}$ (see~\cite{Gralla:2017djj} for a similar discussion in terms of the gravitational waveform). A particularly suitable scenario is the binding energy of a large body tidally deformed by a small perturber~\cite{LeTiec:2011dp,Bini:2013rfa,Kavanagh:2015lva} computed within the self-force approach~\cite{Poisson:2011nh,Sasaki:2003xr}. However, as mentioned above, at a fixed multipole order $\ell \in \mathbb{N}$, $\ell>1$ in four spacetime dimensions $\hat{d} = 1$, knowledge of the (generally unknown) nontidal terms at $1+2\ell$ PN order is required to discriminate the tidal contribution.

While recent progress in the PN expansion~\cite{Blumlein:2020pyo} on $c^{\rm PN}_5$ makes such a matching in four spacetime dimensions feasible for the leading order tidal interaction ($\ell=2$) in the near future, the higher multipole orders would still require a different approach. Alternatively,  analytically continuing self-force results in the dimension or multipole will manifestly separate tidal terms from nontidal PN terms, which need not be known explicitly in this case. The utility of analytic continuation, usually in dimension, to facilitate the matching of effective theories  by avoiding high-order computations in the effective theory has long been established~\cite{Georgi:1993mps}, and was discussed in the context of tidal coefficients in~\cite{Kol:2011vg}. 

This advantageous feature of analytic continuation, combined with the use of (gauge-invariant) observables to extract the tidal information, continues as a main theme in the remainder of this paper. However, instead of working with the binding energy or waveform of a binary system, we consider instead the scattering of dynamical tidal fields off a single compact object~\cite{Bernuzzi:2008rq,Chakrabarti:2013lua,Bautista:2021wfy} and the corresponding observables below.

We also note that a small caveat in the distinction between tidal and PN terms could arise when tidal terms in the effective theory play the role of counterterms which cancel divergences in the nontidal PN terms, as discussed in~\cite{Kol:2011vg}. This would inevitably entangle the tidal and nontidal contributions to (finite) observables, and in particular lead to scale-dependent tidal coefficients. However, such a mild form of ambiguity is a well- understood issue in similar contexts in particle physics. That is, in a given regularization and renormalization scheme, the (finite part of the) Love number can still be uniquely matched using analytic continuation arguments~\cite{Kol:2011vg}, including its scale dependence. These subtleties are absent for adiabatic tides in four spacetime dimensions~\cite{Kol:2011vg} but play a role for dynamical tides, as seen e.g. in the scale-dependence of dynamical quadrupoles obtained in~\cite{Goldberger:2009qd}.

Finally, we comment on slightly different definitions (or understanding) of the tidal coefficients (or Love numbers) used in the literature in relation to the convention adopted here. As explained in the introduction, the Love number can be understood as the ratio of coefficients in the metric with distinct static asymptotic behavior~\cite{Hinderer:2007mb,Binnington:2009bb}, hinging on a choice of coordinates. Alternatively, the Love number or more generally the dynamical tidal response can be seen as a property of an effective theory for a compact object such as a coupling constant in an effective action~\cite{Goldberger:2005cd,Damour:2009vw,Kol:2011vg}, which is the convention adopted here. Finally, one may define the Love number based on observables, in particular relative to possible (conservative) tidal effects of black holes as a baseline~\cite{Gralla:2017djj}, which ultimately seems most advantageous as unlike the other approaches, it does not suffer from either a coordinate dependence or a possible scale (and renormalization-scheme) dependence. Fortunately, all approaches are equivalent in four spacetime dimensions for the adiabatic tidal Love numbers: the connection of coefficients in the asymptotic expansion of the metric to the effective action was derived in~\cite{Kol:2011vg}, which also provided a proof that the Love numbers of nonspinning black holes vanish, and we recapitulated above the straightforward connection between tidal contributions to observables and the effective theory.


\section{Love numbers from scattering}
\label{sec:scattering}

Having established a gauge-invariant definition of the adiabatic tidal response in Sec.~\ref{sec:bindingE} and its connection with quantities appearing in an effective action, let us now work out the case of a fully dynamical tidal response. We focus here on the link between the tidal response in the effective theory and the microphysics of the compact object, which is also advantageous for recovering the adiabatic case. We make this connection by considering wave scattering. For simplicity of developing the general methodology for this scenario, we consider scalar waves. As before, we work with arbitrary multipole order and dimensions, however, we do not specialize to the adiabatic limit of the response. 

\begin{figure}[t]
\begin{center}
		\includegraphics[width=0.48\textwidth,clip]{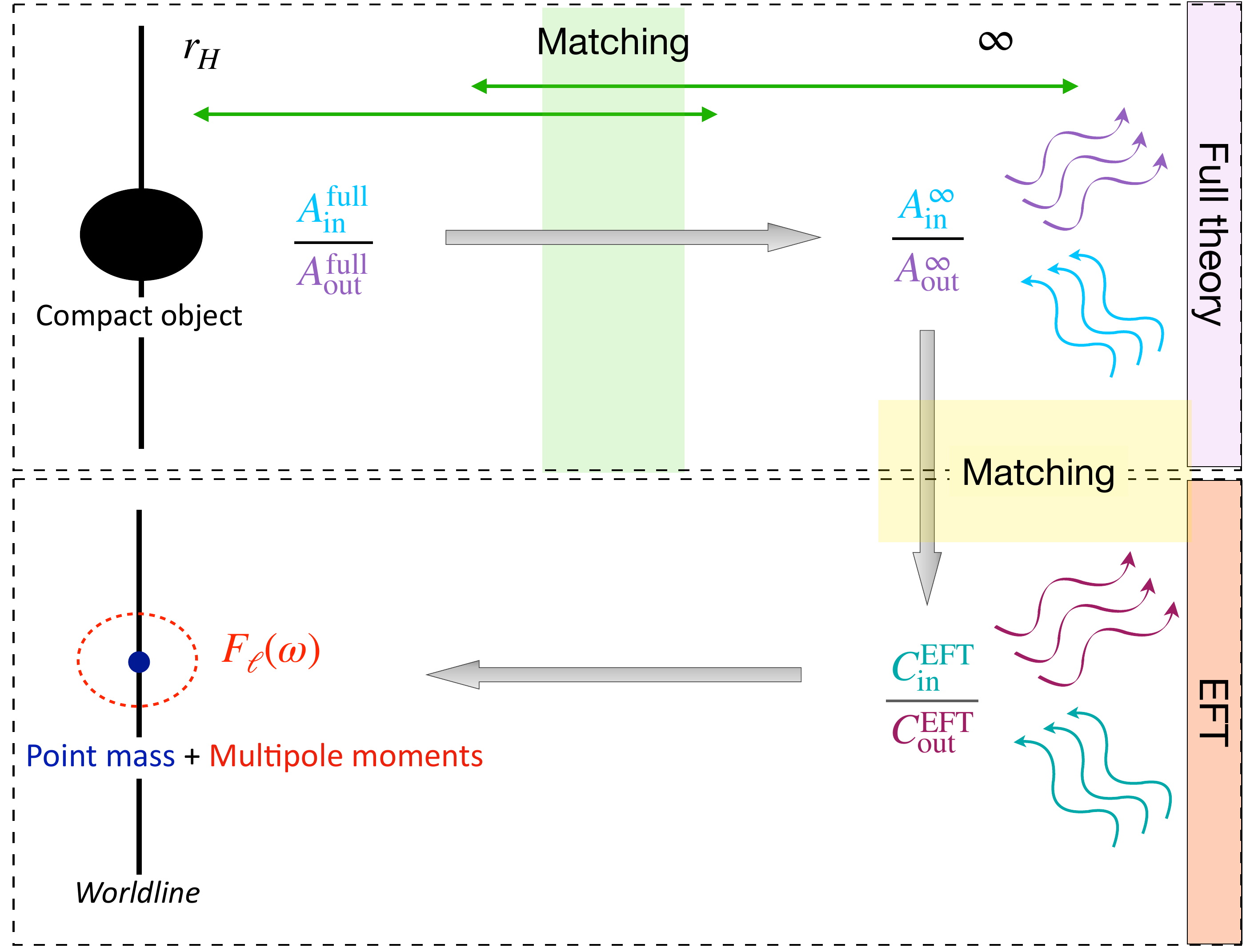}
		\caption{\label{fig:Calculation}
		\emph{Schematic  calculational process to determine the response function $F_\ell(\omega)$}. The response is defined in an effective description (EFT) where the compact object is viewed from large distances and appears as a point mass with multipole moments. Grey arrows indicate the information flow from the microphysical properties of the object to the response via the ratio of in- and outgoing wave amplitudes at infinity. In the full theory calculations based on relativistic perturbations, we specialize to a black hole, thus need only the solutions in the exterior of the horizon $r_H$, and base the matching of the near-horizon and asymptotic solutions on analytical approximations for $M\omega\ll1$. The matching to the effective theory is achieved by identifying in- and outgoing wave states at infinity. }
		\end{center}
\end{figure}

We first consider the effective action for scalar tidal effects~\cite{Kol:2011vg} and delineate the relation between the body's tidal response and the complex amplitudes of the asymptotic in- and outgoing scalar waves. To compute these amplitudes in terms of the detailed properties of the compact object requires going beyond an effective action and considering the full problem of relativistic scalar perturbations to the compact object. As an explicit example, we perform this calculation for a Schwarzschild black hole based on analytical approximations. Finally, we match the information from the full calculation to the effective action description based on light cone coordinates at null infinity and discuss the new insights gained from this approach. Figure~\ref{fig:Calculation} illustrates the information flow we will trace in this section.

Specifically, the action describing the dynamics of the scalar field $\phi$ is given by~\cite{Kol:2011vg}
\be
\label{eq:actionscalar}
S_{\phi}=-\frac{K_{\phi}}{2}\int d^dx \sqrt{-g} \; g^{\alpha\beta}~\nabla_\alpha\phi\nabla_\beta\phi,
\ee
where, $K_\phi$ is a coupling constant and $d$ denotes the number of spacetime dimensions, together with the usual Einstein-Hilbert action for the gravitational field
\be
\label{eq:actionEH}
S_G=\frac{1}{16\pi G_N}\int d^d x\sqrt{-g}R,
\ee
with $R$ the Ricci scalar. The coupling constant $K_\phi$ is defined such that it coincides with the coupling constant of the full theory $S^{\rm full}_{\phi}=-K_{\phi}^{\rm full}/2\int d^dx \sqrt{-g} g^{\alpha\beta} \nabla_\alpha\phi\nabla_\beta\phi$. Hence $K_\phi=K_{\phi}^{\rm full}$.\footnote{This generalizes the normalization of \cite{Kol:2011vg}. In that work the normalization is chosen to match the scalar field action coming from the Newtonian potential. However,  we consider any type of scalar perturbation and therefore $K_{\phi}^{\rm full}$ will depend on the scalar field producing the tidal perturbation in the full theory. Note that if the scalar field in the full theory is the Newtonian potential, we recover the convention in \cite{Kol:2011vg}.}
We are interested in considering wavelike solutions to the equations of motion derived from this action in two different contexts. In the full problem, we consider the behavior of $\phi$ in the spacetime of a Schwarzschild black hole. This describes linear scalar tidal perturbations of a black hole, since the modification of the spacetime due to the scalar field (i.e.~its energy-momentum tensor) is quadratic in $\phi$. In an effective description, the black hole reduces to a point-mass worldline in flat spacetime with additional nonminimal couplings describing the scalar tidal effects. We show how to extract from these descriptions the tidal response of the black hole based on scattering states defined at null infinity.

\subsection{Effective action for scalar tidal effects and response function in terms of scattering states }
\label{sec:EFT}
In this subsection, we consider a body of mass $m$ perturbed by an external massless scalar field $\phi$. The body responds to the disturbance by developing scalar multipole moments $Q^L$. A similar scenario is also studied in~\cite{Kol:2011vg} and~\cite{Hui:2020xxx}, which specialized to the static response but also included gravitational and vectorial perturbations. 
Here, we are interested in computing the frequency-dependent response by considering a scattering process, where $\phi$ describes in- and outgoing scalar waves. 
As the in- and outgoing states are defined asymptotically at null infinity, it is appropriate to formulate the effective action describing the process in flat spacetime. We first calculate the identification between the induced moments $Q^L$ and the amplitudes of the in- and outgoing wave states. We then compute the response function $F_\ell(\omega)$ characterizing the ratio between the induced tidal moments $Q^L$ and the strength of the tidal perturbation $E^L$. Specifically, the response function $F_\ell(\omega)$ is defined by
\begin{align}
\label{eq:responsedef}
Q^L(\omega)=-F_\ell(\omega)E^L(\omega).
\end{align}
Intuitively, and in analogy with the Newtonian gravitational definitions of the tidal field in~\eqref{eq:Uextexpand}, the externally sourced tidal field $E_L$ corresponds to moments of $\phi$ that are nonsingular at the worldline of the compact object $r=0$. Mathematically, this leads to the
scalar tidal tensor in~\eqref{eq:responsedef}  given by 
\be 
\label{eq:ELdef}
E_L=\underset{r\rightarrow{0}}{\text{FP}}\p_L\phi(\omega).
\ee 
Here, $\underset{r\rightarrow{0}}{\text{FP}}$ denotes the finite part as $r\rightarrow{0}$, which we understand here simply as its value in dimensional regularization, see Appendix~\ref{app:AppendixFP}.\footnote{This essentially corresponds to making use of the vanishing of scaleless momentum-space integrals in dimensional regularization. This is not to be confused with Hadamard's \textit{partie finie}.}
 The function $\phi(\omega)$ is the Fourier transform of the scalar wave
\be
\label{eq:phiofomega}
\phi(\omega)=\frac{1}{\sqrt{2\pi}}\int dt e^{-i\omega t} \phi(t).
\ee
In general, the response function $F_\ell(\omega)$ in~\eqref{eq:responsedef} is complex. In the limit $\omega\to 0$, the $O(\omega^0)$ term is real and  $F_\ell(\omega)$ reduces to the static tidal deformability parameter as the scalar analog of~\eqref{eq:QLlambda}. 
By contrast, the next-order $O(\omega^1)$ term is imaginary and describes dissipation.

The aim of this subsection is to arrive at an expression for the response function in terms of the in- and outgoing wave amplitudes. These amplitudes encode information on the microphysics of the body $m$, which we will discuss in detail in subsequent sections.

\subsubsection{Effective action and equations of motion}
An effective action provides a useful description at large distances from a stellar object. In this regime, the object can be described as a point-particle reference worldline with additional couplings related to tidal effects, similar to the considerations in Sec.~\ref{sec:bindingE}. The effective action can be written as
\begin{align}
S=S_{\rm pm}+S_{\rm tidal}+S_\text{int}+S_{\phi}+ S_G, \label{eq:actionEFT}
 \end{align}
 where $S_\phi$ and $S_G$ are given in~\eqref{eq:actionscalar} and ~\eqref{eq:actionEH} above. The point-mass (pm) action is given by
\be
S_{\rm pm}=-m\int d\tau\sqrt{-u_\mu{u^\mu}},
\ee
where $\tau$ is an affine parameter and $u^\mu=dz^\mu/d\tau$ is the tangent to the worldline $z^\mu(\tau)$.

The Lagrangian for the scalar tidal couplings between tidal moments of the external field $\nabla_L\phi$ and the body's multipole moments is given by
\be\label{eq:actiontidal}
S_{\rm tidal}=-K_{\rm Q}\int d\tau\sqrt{-u_\mu{u^\mu}}\sum_{\ell=0}^\infty{\frac{1}{\ell!}Q^L\nabla_L\phi}.
\ee
Here, $Q^L$ are the multipole moments of the body  and $K_{\rm Q}$ is the coupling constant. 
Finally, the action $S_\text{int}$ describes the internal dynamics, that is the dynamics of the multipoles, see~\eqref{eq:Lint} for a simple example.
We are not specifying it here explicitly, but rather write the solution to the equations of motion for the multipoles in terms of the response function~\eqref{eq:responsedef}.
This equivalently captures the internal dynamics and more naturally covers the case of tidal dissipation, as opposed to $S_\text{int}$. For a detailed derivation we refer to \cite{Chakrabarti:2013xza}.

The equation of motion for the scalar field derived from the action~\eqref{eq:actionEFT} is a sourced wave equation
\begin{align}
\label{eq:eomphi}
\nabla_\mu\nabla^\mu\phi=T_{\phi},
\end{align}
with $T_{\phi}=-(\delta S_{\rm tidal}/\delta\phi)/(K_\phi \sqrt{-g})$ given by
\be
T_{\phi}=\frac{K_Q}{K_{\phi}}\int d\tau \sum_{\ell=0}^{\infty}\frac{(-1)^\ell}{\ell!}\nabla_L\left[\frac{\sqrt{-u_\mu{u^\mu}}}{\sqrt{-g}}Q^L\delta^{(d)}(x^\mu-z^\mu(\tau))\right].
\ee
Next, we make several specializations. Analogous to the calculation of tidal effects in the binding energy, which only had to be performed to the leading Newtonian order when analytic continuation is employed, it is sufficient here to work at linear order in a weak field expansion and disregard the gravitational interaction between the point-mass and the scalar field. That is, we work in flat spacetime, where $\sqrt{-g}=1$ and the covariant derivatives reduce to partial derivatives. We also specialize to the rest frame where $u^\mu=(1,0,0,0)$, and we can set $z^\mu=(\tau, 0, 0 ,0)$ by re-parameterization invariance. With these choices,~\eqref{eq:eomphi} becomes 
\begin{align}
\p_\mu\p^\mu\phi&=\frac{K_Q}{K_{\phi}}\int d\tau \sum_{\ell=0}^{\infty}\frac{(-1)^\ell}{\ell!}\p_L\left[Q^L(\tau)\delta^{(d)}(x^\mu-z^\mu(\tau))\right]\nonumber\\
&= \frac{K_Q}{K_{\phi}}\sum_{\ell=0}^{\infty}\frac{(-1)^\ell}{\ell!}\p_L\left[Q^L(t)\delta^{(D)}(x^i)\right].\label{eq:waveeqTD}
\end{align}
where we used that\footnote{From now on we omit the superindex $D$ of the Dirac delta.}
\be
\delta^{(d)}(x^\mu-z^\mu(\tau))=\delta(t-z^0)\delta^{(D)}(x^i-z^i).
\ee
Taking the Fourier transform of the right-hand side of~\eqref{eq:waveeqTD} with the conventions as in~\eqref{eq:phiofomega} and looking at a fixed frequency leads to
\be
 \p_\mu\p^\mu\phi=\frac{K_Q}{K_{\phi}}\sum_{\ell=0}^{\infty}\frac{(-1)^\ell}{\ell!{\sqrt{2\pi}}}Q^L(\omega)e^{i\omega{t}}\p_L\delta(x^i). \label{eq:waveequation}
\ee

Note that in the static case \eqref{eq:waveequation} reads
\be
\nabla^2\phi=\frac{K_Q}{K_{\phi}}\sum_{\ell=0}^{\infty}\frac{(-1)^\ell}{\ell!}Q^L\p_L\delta(x^i)~,
\ee
and has the solution
\be
\phi=\frac{K_Q}{\Omega_{\dhat+1}K_\phi}\sum_{\mathcal{\ell}=0}^{\infty} \frac{1}{\ell!}\frac{(2\ell+\dhat-2)!!}{\dhat!!}\frac{Q_Ln^{L}}{r^{\ell+d}}~,
\ee
with $\Omega_{\dhat+1}=2\pi^{\dhat/2+1}/\Gamma(\dhat/2+1)$ the volume of the $\dhat+1$-hypersphere. The prefactor is important when computing the zero-frequency Love number. The coupling constant will modify the decaying part of the full scalar field solution and therefore has to be taken into account when computing the multipole moments as the decaying part of the scalar perturbation solution at infinity. In the $\dhat=1$ Newtonian case this subtlety is not present because $K_Q=1$ and $K_\phi=1/\Omega_{2}$.

Next, we solve for the multipole moments $Q^L$ in terms of properties of the in- and outgoing waves of the scattering process. The idea is to explicitly construct the scattering states, then substitute these solutions into the left-hand side of~\eqref{eq:waveequation}. Upon applying the wave operator to the solution, only the components of the scattered waves that depend on the induced multipole moments will contribute a source term, since the scalar tidal field is an external, sourcefree field. This enables us to identify the resulting source terms with the right-hand side of~\eqref{eq:waveequation} and read off the moments $Q_L$. 
\subsubsection{In- and outgoing wave solutions}\label{ssec:inoutwavesols}
For simplicity, we start by considering the solutions with $\ell=0$ and subsequently generate the solution for arbitrary multipoles by applying $\ell$ STF derivatives.
For $\ell=0$, the source term in~\eqref{eq:waveequation} vanishes. In addition, the d'Alembertian operator becomes independent of the angular variables. The equation of motion of the scalar field~\eqref{eq:waveequation} then reduces to
\begin{align}
-\p_t^2\phi^{(0)}+\frac{1}{r^{\dhat+1}}\p_r\left(r^{\dhat+1}\p_r\phi^{(0)}\right)=0,
\end{align}
where $\phi^{(0)}$ is the field with $\ell=0$. Upon decomposing the field as  $\phi^{(0)}= r^{-\dhat/2}f_{\omega}(r)e^{i\omega t}$, this turns into a Bessel-type differential equation,
\begin{align}
4r^2\p_r^2f_{\omega}(r)+4r\p_rf_{\omega}(r)+r^{\dhat/2-1}\left[\omega^2r^2-\frac{\dhat^2}{4}\right]f_{\omega}(r)=0.
\end{align}
The solution can be constructed using Hankel functions:
\begin{align}\label{eq:Hankelphi}
\frac{\phi^{(0)}}{e^{i\omega{t}}}=C_1r^{-\dhat/2}H_{\dhat/2}^{(1)}(\omega{r})+C_2r^{-\dhat/2}H_{\dhat/2}^{(2)}(\omega{r}),
\end{align}
with $C_1$ and $C_2$ constants that are determined by boundary conditions. This solution can be understood as an outgoing and incoming wave, i.e.
\begin{subequations}\label{eq:inoutinc1c2}
\begin{align}
{\phi}^{(0)}_{\rm in}&=C_1r^{-\dhat/2}H_{\dhat/2}^{(1)}(\omega{r})e^{i\omega{t}}~,\\
{\phi}^{(0)}_{\rm out}&=C_2r^{-\dhat/2}H_{\dhat/2}^{(2)}(\omega{r})e^{i\omega{t}}~.
\end{align}
\end{subequations}
To determine the constants $C_1$ and $C_2$ we consider the asymptotic behavior $r\rightarrow\infty$ 
and fix it such that  
\begin{align}
\label{eq:cinoutdef}
{\lim_{r\to\infty}}\color{black}\phi^{(0)}_{\rm in,\,out}\equiv C_{\rm in ,\,out}\frac{e^{i\omega(t\pm{r})}}{r^{\frac{1}{2}(\dhat+1)}}~.
\end{align}
Using the properties of the Bessel functions in~\eqref{eq:inoutinc1c2} we obtain
\be
{\lim_{r\to\infty}}\phi^{(0)}_{\rm in,\, out}=\frac{C_{1,\, 2}}{r^{(\dhat+1)/2}}e^{i\omega{t}}\sqrt{\frac{2}{\pi\omega}}e^{\pm i(\omega{r}-\frac{1}{2}\pi\frac{\dhat}{2}-\frac{1}{4}\pi)}.
\ee
Requiring that this matches~\eqref{eq:cinoutdef}
we see that 
\be
C_{1,\,2}=C_{\rm in,\, out}\sqrt{\frac{\pi\omega}{2}}~e^{\pm i \frac{\pi}{4}(\dhat+1)}.
\ee

As mentioned above, our aim is to relate the source term of~\eqref{eq:waveequation} to the source term obtained from the scattering waves solution~\eqref{eq:inoutinc1c2}, from which we can then determine $Q_L$ in terms of $C_{\rm in,\, out}$. The in- and outgoing basis adapted to the physical states is, however, inconvenient for achieving such an identification directly. It is simpler to use a basis adapted to the different analytical behaviors of the solution, and relate the results to the in- and outgoing states at the end of the calculation. The external field corresponds to a sourcefree solution that is everywhere regular and in particular finite near the origin corresponding to the body's worldline. The contribution from the body's response captures the source of the full solution and diverges near the worldline, corresponding to an irregular solution that is singular at the origin. Hence, working in the basis of regular and irregular solutions disentangles the contributions and corresponding source terms, similar to the methods for identifying the Coulomb field of a body~\cite{Barack:2018yvs}.

\subsubsection{Change of basis}

The basis of regular and irregular solutions is obtained from the in- and outgoing solutions by going from the Hankel functions to the Bessel functions of the first and second kind defined by
\begin{subequations}\label{eq:HankelBesselrelation}
\begin{align}
H_{\dhat/2}^{(1)}(\omega{r})&=J_{\dhat/2}(\omega{r})+iY_{\dhat/2}(\omega{r}),\\
H_{\dhat/2}^{(2)}(\omega{r})&=J_{\dhat/2}(\omega{r})-iY_{\dhat/2}(\omega{r}).
\end{align}
\end{subequations}
Inserting this into~\eqref{eq:Hankelphi} we obtain
\be
\phi^{(0)}=\phi^{(0)}_{\rm reg}+\phi^{(0)}_{\rm irreg}, 
\ee
where the regular and irregular solutions are given by
\begin{subequations}
\label{eq:phiregirreg}
\begin{eqnarray}
\phi^{(0)}_{\rm reg}&=&C_{\rm reg}e^{i\omega{t}}\sqrt{2\pi\omega}~r^{-\dhat/2}J_{\dhat/2}(\omega{r}),\\
\phi^{(0)}_{\rm irreg}&=&C_{\rm irreg}e^{i\omega{t}}\sqrt{2\pi\omega}~r^{-\dhat/2}Y_{\dhat/2}(\omega{r}),
\end{eqnarray}
\end{subequations}
and the coefficients $C_{\rm reg/irreg}$ are related to the constants $C_{\rm in/out}$ by
\begin{subequations}\label{eq:constantsl0}
\begin{align}
C_{\rm reg}&=\frac{\left(C_1+C_2\right)}{\sqrt{2\pi\omega}}\nonumber\\&=\frac{1}{2}\left(C_{\rm out}e^{i\frac{\pi}{4}(\dhat+1)}+C_{\rm in}e^{-i\frac{\pi}{4}(\dhat+1)}\right),\\
C_{\rm irreg}&=\frac{i\left(C_1-C_2\right)}{\sqrt{2\pi\omega}}\nonumber\\&=i\frac{1}{2}\left(C_{\rm out}e^{i\frac{\pi}{4}(\dhat+1)}-C_{\rm in}e^{-i\frac{\pi}{4}(\dhat+1)}\right).
\end{align}
\end{subequations}
\subsubsection{Angular dependence}

Having obtained the solutions for the scalar field for $\ell=0$ we will next apply partial STF derivatives, corresponding to a spherical-harmonic decomposition. The goal is to recover the angular dependence from the $\ell=0$ solution and relate the physical amplitudes $C^L_{\rm in/out}$ to the regular/irregular basis $C^L_{\rm reg/irreg}$ for generic multipolar order $\ell$.
Therefore, the full solutions for arbitrary multipolar order $\ell$ are given by 
\begin{align}
\label{eq:varphi}
\phi
=\sum_{\ell=0}^\infty\left({C}_{\rm reg}^L\p_L\phi^{(0)}_{\rm reg}+{C}_{\rm irreg}^L\p_L\phi^{(0)}_{\rm irreg}\right),
\end{align}
where $\phi^{(0)}_{\rm reg/irreg}$ are given in~\eqref{eq:phiregirreg} and we absorb the constants $C_{\rm reg/irreg}$ in the coefficients $C^L_{\rm reg/irreg}$. To obtain the coefficients $C^L_{\rm in/out}$ in terms of the $\ell=0$ amplitudes~\eqref{eq:constantsl0} we compute the STF derivatives explicitly. 
We use the relation~\cite{Blanchet:1985sp}
\begin{align}
\p_Lf(r)=n_Lr^\ell\left(\frac{1}{r}\frac{\p}{\p{r}}\right)^\ell{f(r)},
\end{align}
and the property for a generic Bessel function $\mathcal{B}_\nu(z)$ of degree $\nu$~\cite{NIST:DLMF}
\begin{align}
&\left(\frac{1}{z}\frac{d}{d{z}}\right)^k\left(z^{\nu}\mathcal{B}_\nu(z)\right)=z^{\nu-k}\mathcal{B}_{\nu-k}(z),\\
&\left(\frac{1}{z}\frac{d}{d{z}}\right)^k\left(z^{-\nu}\mathcal{B}_\nu(z)\right)=(-1)^kz^{-\nu-k}\mathcal{B}_{\nu+k}(z).
\end{align}
Thus, 
\begin{align}
\p_L\left(r^{-\dhat/2}\mathcal{B}_{\dhat/2}(\omega{r})\right)
&=(-1)^\ell{n_L}r^{-\dhat/2}\omega^\ell \mathcal{B}_{\dhat/2+\ell}(\omega{r}).
\end{align}
Putting all together we have  
\begin{align}\label{eq:ldependentasymptsolution}
\phi=\sum_{\ell=0}^\infty& e^{i\omega{t}}\sqrt{2\pi\omega}~r^{-\dhat/2}\omega^\ell{n_L}(-1)^\ell\nonumber\\&\times\left({C}_{\rm reg}^LJ_{\dhat/2+\ell}(\omega{r})+{C}_{\rm irreg}^L Y_{\dhat/2+\ell}(\omega{r})\right).
\end{align}
Recall that this is the solution for a fixed frequency $\omega$ for which $\phi(t,r)=\phi(\omega,r)e^{i\omega t}/\sqrt{2\pi}$, where $\phi(\omega,r)$ coincides with the Fourier transform at a fixed frequency.

We now express the field in the incoming and outgoing basis. In order to do that we obtain the proper asymptotic expression of incoming and outgoing waves for $\ell\neq0$ by proceeding in the same way as above. That is, we apply STF derivatives to the $\ell=0$ expression,\footnote{Note that we are not taking the derivatives on the denominator. This is because asymptotically we do not expect any dependence on the multipole order on the radial denominator. This can also be seen by checking how an angular dependence affects the differential equation for the radial part of the field: the angular eigenvalue $\ell(\ell+1)$ changes the order of the Bessel function but not the factor $r^{\dhat/2}$, which is the responsible term for the numerator. The best example is the wave equation in three spatial dimensions.}
\begin{align}
\label{eq:phiinoutEFTasymptotic}
\lim_{r\rightarrow\infty}\phi_{\rm in/out}&=
C^L_{\rm in/out}\frac{\p_L\left(e^{i\omega(t\pm{r})}\right)}{r^{\frac{1}{2}(\dhat+1)}}\nonumber\\&=C^L_{\rm in/out}n_L(\pm i\omega)^\ell\frac{e^{i\omega(t\pm{r})}}{r^{\frac{1}{2}(\dhat+1)}}.
\end{align}
To obtain the incoming and outgoing solutions we invert Eqs.~\eqref{eq:HankelBesselrelation}, which leads to
\begin{subequations}
	\begin{align}\label{eq:BesselHankelrelation}
	J_{\dhat/2+\ell}(\omega{r})&=\frac{1}{2}\left(H_{\dhat/2+\ell}^{(1)}(\omega{r})+H_{\dhat/2+\ell}^{(2)}(\omega{r})\right),\\
	Y_{\dhat/2+\ell}(\omega{r})&=\frac{1}{2i}\left(H_{\dhat/2+\ell}^{(1)}(\omega{r})-H_{\dhat/2+\ell}^{(2)}(\omega{r})\right),
	\end{align}
\end{subequations}
and identify the incoming and outgoing solutions with the first- and second-order Hankel functions, respectively,
 \begin{align}
 \phi_{\rm in}=&\sum_{\ell=0}^\infty e^{i\omega{t}}\sqrt{2\pi\omega}~r^{-\dhat/2}\omega^\ell{n_L}\frac{(-1)^\ell}{2} H_{\dhat/2+\ell}^{(1)}(\omega{r})\nonumber\\&\times\left({C}_{\rm reg}^L-i{C}^L_{\rm irreg}\right),\\
 \phi_{\rm out}=&\sum_{\ell=0}^\infty e^{i\omega{t}}\sqrt{2\pi\omega}~r^{-\dhat/2}\omega^\ell{n_L}\frac{(-1)^\ell}{2} H_{\dhat/2+\ell}^{(2)}(\omega{r})\nonumber\\&\times\left({C}^L_{\rm reg}+i{C}^L_{\rm irreg}\right).
 \end{align}
We use the asymptotic behavior of the Hankel functions~\cite{NIST:DLMF}
\begin{align}
H_{\nu}(z)\sim\sqrt{\frac{2}{\pi z}}e^{\pm{iz}}e^{\mp{i}\frac{\pi}{4}\left(2\nu+1\right)},
\end{align}
where the upper sign applies for $H^{(1)}$ and the lower sign for $H^{(2)}$. With this, we obtain the generalization of~\eqref{eq:constantsl0} for any multipole order $\ell$,
\begin{subequations}\label{eq:Cregiregtoinout}
\begin{align}
C^L_{\rm reg}=\frac{(-1)^\ell}{2}&i^\ell\Big[C^L_{\rm in}e^{i\frac{\pi}{4}(\dhat+2\ell+1)}\nonumber\\&+(-1)^\ell C^L_{\rm out}e^{-i\frac{\pi}{4}(\dhat+2\ell+1)}\Big],\\
C^L_{\rm irreg}=\frac{(-1)^\ell}{2}&i^{\ell+1}\Big[C^L_{\rm in}e^{i\frac{\pi}{4}(\dhat+2\ell+1)}\nonumber\\&+(-1)^{\ell+1} C^L_{\rm out}e^{-i\frac{\pi}{4}(\dhat+2\ell+1)}\Big].
\end{align}
\end{subequations}

\subsubsection{Tidally induced multipoles}

We next compute $Q^L(\omega)$ and its relation to the coefficients $C_{\rm in/out}$ by noting that $Q^L$ can be identified from the source terms in the wave equation, c.f.~\eqref{eq:waveequation}. We can compute this source in terms of $C_{\rm in/out}$ by applying the d'Alembertian to the solutions constructed in the previous subsections. This allows us to read off $Q^L$ in terms of the constants. As above, for convenience, we first work in the regular/irregular basis and transform to the in/out basis at the end, and also first consider $\ell=0$, then generate the angular dependence through STF derivatives.


When applying operators to the solution~\eqref{eq:phiregirreg}, they must be understood in a distributional sense. The reason is that as the equation of motion~\eqref{eq:waveeqTD} indicates, the source is only defined in a distributional manner. We will denote the distributional operators with a tilde, e.g. $\tilde \nabla^2 $ is the distributional Laplace operator.

We first consider the distributional Laplacian of the regular solution in~\eqref{eq:phiregirreg}
using the series representation of the Bessel functions 
around $r=0$ given by~\cite{AbraStegun}
\begin{align}
\label{eq:Jseries}
J_\nu(\omega{r})=\sum_{k=0}^\infty\frac{(-1)^k}{k!\Gamma(k+\nu+1)}\left(\frac{\omega{r}}{2}\right)^{2k+\nu}~,
\end{align}
where the sum is over positive integers $k\in \mathbb{Z}^+$. Inserting~\eqref{eq:Jseries}
with $\nu=\dhat/2$ in~\eqref{eq:phiregirreg} we obtain
\begin{align}\label{eq:nabla2reg}
\frac{\tilde{\nabla}^2\phi^{(0)}_{\rm reg}}{C_{\rm reg}e^{i\omega{t}}\sqrt{2\pi\omega}}
&=\sum_{k=0}^\infty h_k^+\tilde{\nabla}^2\left(r^{2k}\right)\nonumber\\
&=-r^{-d/2}\omega^2J_{\dhat/2}(\omega{r}),
\end{align}
where 
\be
\label{eq:hkdef}
h_k^\pm(\omega)=\frac{(-1)^k}{k!\Gamma(k\pm\frac{\dhat}{2}+1)}\left(\frac{\omega}{2}\right)^{2k\pm \dhat/2}.
\ee
Here, we used the results from Appendix~\ref{app: DistrLaplacian} for the distributional Laplacian acting on $r^{-\beta}$ for any $\beta \in \mathbb{R}$ 
\be
\label{eq:distrLaplmain}
\tilde{\nabla}^2\left(\frac{1}{r^\beta}\right)=\begin{cases}{\nabla}^2r^{-\beta}-\dfrac{2\,\hat d\,\pi^{1+\hat d/2}}{\Gamma\left(1+\frac{\hat d}{2}\right)}\delta(x^i), \;\; \beta=\hat d\in\mathbb{Z}\\
{\nabla}^2r^{-\beta} \qquad \qquad \qquad \beta<\hat d
\end{cases},
\ee
with $\beta=-2k<\dhat$. We then used the identity for the standard Laplacian
\begin{align}
\nabla^2{r^\beta}=\beta(\beta+\dhat)r^{\beta-2}, 
\end{align}
with $\beta=2k$ and resummed the series into the Bessel function as per~\eqref{eq:Jseries}.

For the Laplacian of the irregular solution~\eqref{eq:phiregirreg} we first work with odd values of $\dhat$ and take the limit for even values at the end using L'H\^{o}pital's rule~\cite{Arfken}. The Bessel function of the second kind for odd $\dhat$ reads~\cite{AbraStegun}
\begin{align}\label{eq:Besselsecondkindintermsoffirst}
Y_{\dhat/2}(\omega{r})=\frac{1}{\sin\left(\frac{\pi\dhat}{2}\right)}\left[\cos\left(\frac{\pi\dhat}{2}\right)J_{\dhat/2}(\omega{r})-J_{-\dhat/2}(\omega{r})\right]~.
\end{align}
Applying the Laplacian to the irregular solution~\eqref{eq:phiregirreg} and using~\eqref{eq:Besselsecondkindintermsoffirst} yields
\begin{align}
\label{eq:Laplirreg1}
&\frac{\sin\left(\frac{\pi\dhat}{2}\right)\tilde{\nabla}^2\phi^{(0)}_{\rm irreg}}{C_{\rm irreg}e^{i\omega t} \sqrt{2\pi\omega}}=\sin\left(\frac{\pi\dhat}{2}\right)\tilde{\nabla}^2\left[r^{-\dhat/2}Y_{\dhat/2}(\omega{r})\right]\nonumber\\
&\quad =-\cos\left(\frac{\pi\dhat}{2}\right)r^{-d/2}\omega^2J_{\dhat/2}(\omega{r}) -{\cal S},
\end{align}
where we used~\eqref{eq:nabla2reg} for the first term, and used the series expansion~\eqref{eq:Jseries} to define
\be
{\cal S}=\sum_{k=0}^\infty h_k^-(\omega)\tilde{\nabla}^2\left(r^{2k-\dhat}\right), \label{eq:Sdef}
\ee
where $h_k^-$ is given in~\eqref{eq:hkdef}.
We now compute explicit results for ${\cal S}$.
Since the dimension is an arbitrary parameter $\dhat\geq1$, we split the series into a contribution from positive and negative powers of $r$ corresponding to $k> \lfloor\hat d/2\rfloor$ and $k <\lfloor \hat d/2\rfloor$ respectively in~\eqref{eq:Sdef}. Here, $\lfloor \ldots \rfloor$ denotes the floor function. For the positive powers of $r$, the action of the distributional Laplacian is the same as the usual Laplacian. This follows from the second case in~\eqref{eq:distrLaplmain} with $\beta=\dhat-2k$, which for $k> \hat d/2$ is always $ \beta<\dhat$. For the series involving negative powers of $r$ we use~\eqref{eq:distrLaplmain} with $\beta=2k-\dhat$ and the index $k$ running from $k=0$ to $k<\lfloor \hat d/2\rfloor$. We see that singular contributions involving the Dirac-$\delta$ only arise when $k=0$.
The remaining terms from~\eqref{eq:distrLaplmain} involving the standard Laplacian in the series recombine with that from the positive powers of $r$ into a single series over all $k$. Altogether, this leads to
\begin{align}
\label{eq:calSresult}
&{\cal S}=\sum_{k=0}^\infty h_k^-(\omega){\nabla}^2\left(r^{2k-\dhat}\right)\nonumber\\
&\quad -\sum_{k=0}^{\lfloor \dhat/ 2 \rfloor}h_k^-(\omega) \frac{2\, \dhat\, \pi^{1+\hat d/2}}{\Gamma\left(1+\frac{\hat d}{2}\right)}\delta_{k,0}\delta(x^i)\nonumber\\
=&-r^{-d/2}\omega^2 J_{-\dhat/2}(\omega{r})-\left(\frac{\omega}{2}\right)^{-\dhat/2}4\pi^{\dhat/2}\sin\left(\pi\frac{\dhat}{2}\right)\delta(x^i),
\end{align}
 where in the first equality the Kronecker delta $\delta_{k,0}$ accounts for the fact that the only nonzero contribution involving $\delta(x^i)$
 arises from $k=0$. In the last line of~\eqref{eq:calSresult} we used
\begin{align}
\Gamma\left(1+\frac{\dhat}{2}\right)\Gamma\left(1-\frac{\dhat}{2}\right)=\frac{1}{\sin\left(\pi\frac{\dhat}{2}\right)}\frac{\pi\dhat}{2}~.
\end{align}
Inserting the result~\eqref{eq:calSresult} into the Laplacian of the irregular solution~\eqref{eq:Laplirreg1} leads to
\begin{align}
\label{eq:nabla2irreg}
\frac{\tilde{\nabla}^2\phi^{(0)}_{\rm irreg}}{C_{\rm irreg}e^{i\omega{t}}\sqrt{2\pi\omega}}=\left(\frac{\omega}{2}\right)^{-\dhat/2}4\pi^{\dhat/2}\delta(x^i)-\omega^2r^{-\dhat/2} Y_{\dhat/2}(\omega{r}) .
\end{align}
From these results for the Laplacian of the solutions, we finally compute the action of the d'Alembertian  $\tilde{\Box}=\tilde\partial_\mu \tilde \partial^\mu=-\partial_t^2+\tilde{\nabla}^2$. The time dependencies of the fields only enter through $e^{i\omega t}$, and thus, the term involving second time derivatives in $\tilde \Box$ will lead to $\omega^2\phi^{(0)}_{\rm reg/irreg}$. This cancels with those terms coming from the action of the Laplacian that are directly proportional to the Bessel functions in Eqs.~\eqref{eq:nabla2reg} and~\eqref{eq:nabla2irreg}. Consequently, upon applying the d'Alembertian to the solution all terms proportional to a Bessel function will vanish, and we obtain
\begin{subequations}
\label{eq:Boxofreg}
\begin{align}
\tilde{\Box}\phi^{(0)}_{\rm reg}&=0~,\\
\tilde{\Box}\phi^{(0)}_{\rm irreg}&=C_{\rm irreg} e^{i\omega{t}}8\pi\left(\frac{2\pi}{\omega}\right)^{\frac{\dhat-1}{2}}\delta(x^i)~.\label{eq:singularpartwaveeq}
\end{align}
\end{subequations}
We see that the source term corresponding to the irregular solution is non-singular for both odd and even values of $\dhat$.

Having worked out the results~\eqref{eq:Boxofreg} for $\ell=0$, the final step is to obtain the angular dependencies for arbitrary multipole moments. In order to compute the d'Alembertian of the solution for generic multipolar order we will apply the same strategy as above. This is, we will apply STF derivatives to the $\ell=0$ d'Alembertian and use the commutativity of both operators\footnote{The commutativity of distributional derivatives can readily be seen in the Fourier domain, where they correspond to a multiplication by the wave vector.},
\begin{align}
\tilde{\Box}\phi&=\tilde{\Box}\left({C}_{\rm irreg}^L\p_L\phi^{(0)}_{\rm irreg}+{C}_{\rm reg}^L\p_L\phi^{(0)}_{\rm reg}\right)\nonumber\\
&={C}_{\rm irreg}^Le^{i\omega{t}}8\pi\left(\frac{2\pi}{\omega}\right)^{\frac{\dhat-1}{2}}\p_L\delta(x^i)\label{eq:finalBoxphi}.
\end{align}

Next, we use this result to identify how the scalar multipole moments $Q^L$ are encoded in the coefficients ${C}^L_{\rm reg/ irreg}$. 
Comparing~\eqref{eq:finalBoxphi} with the wave equation~\eqref{eq:waveequation}, we infer
\begin{align}
\label{eq:QLresult}
Q^L(\omega)=\frac{K_\phi}{K_{Q}}\ell!(-1)^{\ell}{8\pi\sqrt{2\pi}}\left(\frac{2\pi}{\omega}\right)^{\frac{\dhat-1}{2}}{C}_{\rm irreg}^L~.
\end{align}

\subsubsection{The response function and its relation to in- and outgoing wave amplitudes}

Let us come back to the calculation of the response function defined in~\eqref{eq:responsedef}.
With an expression for the tidally induced multipoles at hand~\eqref{eq:QLresult}, we are missing an expression for the finite part of the STF derivatives of $\phi$. Hence we first compute 
\begin{align}\label{eq:PLphisol}
\p_L\phi=\sum_{k=0}^\infty\left({C}_{\rm reg}^K\p_L\p_K\phi^{(0)}_{\rm reg}+{C}_{\rm irreg}^K\p_L\p_K\phi^{(0)}_{\rm irreg}\right)~.
\end{align}
In order to extract the finite part we directly substitute the series representation and apply the STF derivatives to the regular/irregular part. We refer to Appendix \ref{app:AppendixFP} for the details of the computation. We obtain that the finite part of the field determining the tidal tensor defined in~\eqref{eq:ELdef} is
\be
\label{eq:FPdLphi}
E_L(\omega)=e^{i\omega{t}}\ell!\pi\left(\frac{\omega}{2}\right)^{\dhat/2+1/2+2\ell}\frac{(-1)^{\ell}2^{\ell+1}}{\Gamma(\frac{\dhat}{2}+\ell+1)}{C}_{\rm reg}^L~.
\ee
where we use that $\phi(\omega)=\sqrt{2\pi}e^{-i\omega t}\phi(t)$ for a fixed frequency $\omega$.
With the results of~\eqref{eq:QLresult} and~\eqref{eq:FPdLphi} we can compute the response defined by~\eqref{eq:responsedef}. 
Both the tidal field and the multipoles depend on the tensorial STF coefficients $C^L$. They can be converted to scalar quantities by expressing them in a spherical harmonic basis as discussed in~\cite{Thorne:1980ru}. This decomposition extends to higher dimensions, as can be verified using the hyperspherical harmonics discussed in Sec.\ref{sec:Schwarzschild} and the identities in Appendix~\ref{sec:appendix}, and is given by
\be
\label{eq:STFtospher}
C^L=\sum_mC_{\ell m}\mathcal{Y}^L_{\ell m}.
\ee
Here, $\mathcal{Y}^L_{\ell m}$ are STF tensors with complex coefficients defined by the relation between spherical harmonics $Y_{\ell m}$ and unit vectors through 
\be
\label{eq:Yton}
Y_{\ell m}=\mathcal{Y}^L_{\ell m}n_L.
\ee
Taking into account that spherical symmetry implies that the in/out coefficients $C_{\ell m}$ are independent of the azimuthal number $m$ leads to
\begin{align}
\label{eq:inouttospher}
C^L_{\rm in/out}=C_{\ell}^{ \rm in/out}\sum_m\mathcal{Y}^L_{\ell m},
\end{align}
The ratio needed in the response can thus be expressed as
\begin{align}
\frac{C^L_{\rm in}}{C^L_{\rm out}}=\frac{C_{\ell}^{\rm in}\sum_m\mathcal{Y}^L_{\ell m}}{C_{\ell }^{\rm out}\sum_{m\prime}\mathcal{Y}^L_{\ell {m\prime}}}=\frac{C_{\ell}^{ \rm in}}{C_{\ell}^{ \rm out}}.
\end{align}
 Analogously, by virtue of~\eqref{eq:Cregiregtoinout},
\begin{align}\label{eq:STFtoregirreg}
\frac{C^L_{\rm irreg}}{C^L_{\rm reg}}=\frac{C_{\ell}^ {\rm irreg}}{C_{\ell}^{ {\rm reg}}}~.
\end{align}
From~\eqref{eq:responsedef}, with~\eqref{eq:QLresult} and~\eqref{eq:FPdLphi}, and using \eqref{eq:STFtoregirreg}, we obtain the response function 
\be
F_\ell(\omega)=-\frac{Q^L}{E_L}=
\frac{K_\phi}{K_{Q}}\tilde{F}_{\ell}(\omega),
\ee
 with
\be
\tilde{F}_{\ell}(\omega)=\Xi_\ell\,\frac{C_{\ell }^{{\rm irreg}}}{C_{\ell}^{ {\rm reg}}}
\ee
the normalised response function and
\be
\label{eq:Xildef}
\Xi_\ell=-\frac{4\pi^{\dhat/2}}{2^{\ell}}\left(\frac{2}{\omega}\right)^{\dhat+2\ell}\Gamma\left(\frac{\dhat}{2}+\ell+1\right).
\ee
In particular, $\tilde{F}_{\ell}(\omega)$ will coincide with the definition of the tidal deformability used in \cite{Kol:2011vg}, where $\lambda_\ell$ is independent on the coupling constants. However, this definition differs from \cite{Hui:2020xxx} due to their different normalizations. Additionally, we can also set $K_Q=1$ without loss of generality. This is because when plugging back \eqref{eq:responsedef} into \eqref{eq:actiontidal} we are left with $S_{\rm tidal}\propto K_\phi\int d\tau \sum_{\ell=0} \tilde{F}_\ell E_LE^L/\ell!$ independently of $K_Q$.
Using~\eqref{eq:Cregiregtoinout} we obtain the response function in the in-/outgoing basis 


\be
\label{eq:responsefunction}
\tilde{F}_\ell(\omega)=i\,\Xi_\ell\left[1-\dfrac{2}{1+\frac{C_{\ell}^{ \rm in}}{C_{\ell }^{ \rm out}}e^{i\frac{\pi}{2}\left(\dhat+1\right)}}\right]
\ee
where $\Xi_\ell$ is given in~\eqref{eq:Xildef}. Writing the in-/outgoing complex amplitudes in terms of a complex scattering phase $\delta_\ell$, defined by $C_{\ell}^{\rm in}/C_{\ell}^{\rm out}= e^{2i\delta_\ell}$, we can rewrite \eqref{eq:responsefunction} as
\be
\label{eq:responsefunctionphaseshift}
\tilde{F}_\ell(\omega)=-\Xi_\ell\tan\left[\delta_\ell+\frac{\pi}{4}(\dhat+1)\right]
\ee

Note that in deriving the above results for the response function~\eqref{eq:responsefunction}, there was no need to assume any analytic continuation in $\ell$ or $\dhat$,  except for the fact that the finite part must be obtained  using some regulator.


The values of the coefficients 
$C_{\ell }^{ \rm in/ out}$ depend on the detailed internal structure properties of the body $m$ and cannot be determined within the effective description. Instead, they must be computed from the full description of relativistic perturbations to the compact object under consideration. In the next subsection, we specialize to the body being a nonrotating black hole and perform this perturbation-theory calculation.  

\subsection{Amplitudes of the scattering states for a Schwarzschild black hole}
\label{sec:Schwarzschild}

In general, to determine the detailed information about the compact object contained in the response~\eqref{eq:responsefunction} requires solving for relativistic perturbations in the interior and exterior of the object, matching these solutions, and extracting the asymptotic scattering states. In the special case of black holes, due to the presence of the horizon, the interior calculations are replaced by considering the near-horizon solutions, as we discuss below. The case of a nonspinning black hole is a well-studied example and enables us to check our results from scattering against known results in the literature, namely the static response function~\cite{Kol:2011vg} and the absorption cross section~\cite{Page:1976df}. As we study scalar perturbations, the spacetime remains unaffected and our analysis focuses on the scalar field equations. 

We first calculate the behavior of the field near the horizon. In this regime, there is no closed-form solution to the perturbation equations, though in four spacetime dimensions a highly useful series expansion known as the MST solution~\cite{Mano:1996vt} is available. 
Here, we also make use of analytical approximations valid for $M\omega\ll 1$, where $M$ is the mass of the black hole but work only to the leading order. Next, we consider the perturbation equations in the asymptotic limit of distances much larger than the size of the black hole, $r_H/r\ll 1$, where $r_H$ corresponds to the horizon. These solutions describe waves propagating along the Schwarzschild light cones. The last step in this subsection is to connect the detailed information about the tidal response from the near-horizon regime to the amplitudes of the asymptotic waves, both computed within the relativistic perturbation framework. This is accomplished through matched asymptotic expansions, specifically by considering the near-horizon solutions in the limit $r_H/r\ll 1$ and the asymptotic wave solution for $\omega r\ll1$. We show that these two asymptotic expansions overlap and perform the matching of the coefficients. 

\subsubsection{Scalar wave perturbations to a Schwarzschild black hole}

We start by obtaining the equation of motion of the scalar field on the $\dhat$-dimensional Schwarzschild background using the action for the scalar-field dynamics given in~\eqref{eq:actionscalar}. 
In Schwarzschild coordinates, the spacetime is described by the metric 
\be
\label{eq:Schw}
ds^2=-f(r)dt^2+\frac{1}{f(r)}dr^2+r^2d\Omega_{\dhat+1},
\ee
where $r^2\Omega_{\dhat+1}$ denotes the surface element on a $(\dhat+1)$-dimensional hypersphere
and
\be
\label{eq:fdef}
f(r)=1-\left(\frac{r_H}{r}\right)^{\dhat}.
\ee
Here, $r_H$ is the Schwarzschild radius corresponding to the black hole's horizon. 
Because the spacetime is static and spherically symmetric, we make the following ansatz for the decomposition of the field
\begin{align}
\label{eq:phiSchwans}
\phi=\sum_{\ell m}\int d\omega \frac{e^{i\omega{t}}}{\sqrt{2\pi}}\frac{\psi_{\omega \ell}(r)}{\alpha(r)}Y_{\ell m}(\Omega)~,
\end{align}
where due to spherical symmetry $\psi_{\omega \ell m}(r)=\psi_{\omega \ell}(r)$ and we have introduced 
\be
\label{eq:alphadef}
\alpha(r)=r^{(\dhat+1)/2} ,
\ee
which absorbs the radial dependence of the volume element $\sqrt{-g}\propto r^{\dhat+1}$ into the field. We substitute the ansatz~\eqref{eq:phiSchwans} into the action~\eqref{eq:actionscalar}. For convenience we choose $K_{\phi}=K_{\phi}^{\rm full}=1$. Note that for this choice of coupling constant the response is simply $F_{\ell}(\omega)=\tilde{F}_{\ell}(\omega)$. Using the metric~\eqref{eq:Schw} in \eqref{eq:actionscalar} we obtain
\begin{align}
&S_{\phi}=-\frac{1}{2}\sum_{\ell m}\sum_{\ell' m'}\int d\omega d\omega' \alpha^2 dt dr d\Omega_{\dhat+1}\nonumber\\&\times\Bigg[   f\left(\p_r\frac{\psi^\ast_{\omega' \ell'}}{\alpha}\right)\left(\p_r\frac{\psi_{\omega \ell}}{\alpha}\right)\frac{e^{i (\omega-\omega')t}}{2\pi}Y_{\ell m}Y^\ast_{\ell' m'}-\frac{\omega\omega'}{f}\phi\phi^\ast\nonumber\\&\quad+\frac{e^{i(\omega-\omega')t}}{2\pi\alpha^2}\psi^\ast_{\omega' \ell'}\psi_{\omega \ell}g^{\Omega' \Omega}\nabla_{\Omega'}Y^\ast_{\ell' m'}\nabla_{\Omega}Y_{\ell m}\Bigg]~,
\end{align}
where we used that $\phi$ is a real field such that $\phi(t)=\phi^\ast(t)$. Here, $g^{\Omega\Omega'}=r^{-2}\delta^{\Omega\Omega'}$ and the functions $Y_{\ell m}(\Omega)$ are the hyperspherical harmonics having the properties~\cite{Avery}
\begin{align}\label{eq:hypershpericalharmonics}
\int& d\Omega_{\dhat+1}g^{\Omega' \Omega}\nabla_{\Omega'}Y^\ast_{\ell' m'}\nabla_{\Omega}Y_{\ell m}\nonumber\\&=-\frac{1}{r^2}\int d\Omega_{\dhat+1}Y^\ast_{\ell' m'}\nabla^2Y_{\ell m}\nonumber\\&=\frac{\dhat^2{\lhat(\lhat+1)}}{r^2}\int d\Omega_{\dhat+1}Y^\ast_{\ell' m'}Y_{\ell m}~,
\end{align}
where $\hat{\ell}$ was defined in~\eqref{eq:lhatdef}. Using~\eqref{eq:hypershpericalharmonics} leads to the action

\begin{align}
\label{eq:actionSchwdecomp1}
&S_\phi=-\frac{1}{2}\sum_{\ell m}\sum_{\ell' m'}\int d\omega \int d\omega'\left[\int dt \frac{e^{i(\omega-\omega')t}}{2\pi}\right]\nonumber\\&\times\Bigg\{\int dr \Bigg[\Bigg(-\frac{\omega~\omega'}{f}+\frac{\dhat^2{\lhat(\lhat+1)}}{r^2}\Bigg)\psi^\ast_{\omega' \ell'}\psi_{\omega \ell}\nonumber\\&+f\alpha^2\left(\p_r\frac{\psi^\ast_{\omega' \ell'}}{\alpha}\right)\left(\p_r\frac{\psi_{\omega \ell}}{\alpha}\right)\Bigg]\Bigg\}\left[\int d\Omega_{\dhat+1}Y^\ast_{\ell' m'}Y_{\ell m}\right].
\end{align}
This simplifies upon using the normalizations 
\begin{align}
&\int dt \frac{e^{i(\omega-\omega')t}}{2\pi}=\delta(\omega-\omega')~,\\
&\int d\Omega_{\dhat+1}Y^\ast_{\ell' m'}Y_{\ell m}=\delta_{\ell \ell'}\delta_{m m'}.
\end{align}
Further, the last term inside the curly brackets in~\eqref{eq:actionSchwdecomp1} simplifies when writing out the derivatives, using integration by parts and omitting the total derivative, and can be written as
\begin{align}
\int &dr f\alpha^2\left(\p_r\frac{\psi^\ast_{\omega' \ell'}}{\alpha}\right)\left(\p_r\frac{\psi_{\omega \ell}}{\alpha}\right)\nonumber\\
=&\int dr\left[f\p_r\psi^\ast_{\omega \ell}\p_r\psi_{\omega \ell}+\frac{\p_r\left(f\p_r\alpha\right)}{\alpha}\psi^\ast_{\omega \ell}\psi_{\omega \ell}\right]. 
\end{align}
With these simplifications, the action~\eqref{eq:actionSchwdecomp1} reduces to
\begin{align}\label{eq:Scalaractionnormalcoords}
S_\phi=-\frac{1}{2}&\sum_{\ell m}\int d\omega \int dr\Bigg\{f\p_{r}\psi^\ast_{\omega \ell}\p_{r}\psi_{\omega \ell}\nonumber\\&+\Bigg[\frac{\dhat^2\lhat(\lhat+1)}{r^2}-\frac{\omega^2}{f}+\frac{\p_r\left(f\p_r\alpha\right)}{\alpha}\Bigg]\psi^\ast_{\omega \ell}\psi_{\omega \ell}\Bigg\}~.
\end{align}
The equations of motion derived from this action read
\begin{align}\label{eq:ScalarEoMnormalcoords}
\p_r&\left({f\p_r\psi_{\omega \ell}}\right)\nonumber\\&-\Bigg[\frac{\dhat^2\lhat(\lhat+1)}{r^2}-\frac{\omega^2}{f}+\frac{\p_r\left(f\p_r\alpha\right)}{\alpha}\Bigg]\psi_{\omega \ell}=0~.
\end{align}
It is convenient to transform the radial Schwarzschild coordinate $r$ to the tortoise coordinate $r^\ast$, which is known to lead to the simplest representation of the equations of motion in this problem. The tortoise coordinate is defined by
\begin{align}
\label{eq:rstardef}
dr^\ast=\frac{1}{f(r)}dr,~\quad \p_r=\frac{1}{f(r)}\p_r^\ast. 
\end{align}
In terms of this coordinate, the action~\eqref{eq:Scalaractionnormalcoords} takes the form
\begin{align}
S_\phi=-\frac{1}{2}&\sum_{\ell m}\int d\omega \int dr^\ast[\p_{r^\ast}\psi^\ast_{\omega \ell}\p_{r^\ast}\psi_{\omega \ell}\nonumber\\&+(V_\ell-\omega^2)\psi^\ast_{\omega \ell}\psi_{\omega \ell}]~,
\label{eq:tortoise}
\end{align}
with the potential $V_\ell$ given by
\begin{align}
\label{eq:Vpotential}
V_\ell=&f\left[\frac{\dhat^2\lhat(\lhat+1)}{r^2}+\frac{\p_r\left(f\p_r\alpha\right)}{\alpha}\right]~.
\end{align}
The equation of motion of the scalar field derived from the action~\eqref{eq:tortoise} reads
\begin{align}\label{eq:EoMSchrodinger}
\p^2_{r^\ast}\psi_{\omega \ell}-(V_\ell-\omega^2)\psi_{\omega \ell}=0~. 
\end{align}
This equation has no closed-form analytic solution for generic dimensions and generic frequency. Solutions are only available in the special cases of four spacetime dimensions~\cite{Mano:1996vt} and in the zero-frequency limit~\cite{Kol:2011vg}. For our purposes, it will be sufficient to consider the asymptotic solutions close to the horizon and at large distances from the black hole, as we discuss next. The near-horizon solutions provide the information on the detailed properties of the perturbed black hole in the strong-field regime, while the asymptotic behavior at large distances determines the matching to the effective description of Sec.~\ref{sec:EFT}. This information flow will be traced in detail through the calculations in the next subsections.

\subsubsection{Boundary conditions}
An important preliminary to the analysis of wave solutions is to identify the appropriate boundary conditions. As stressed in~\cite{Chia:2020yla}, the proper treatment of the boundary conditions is crucial in order to unambiguously identify the tidal and multipolar contributions. We start by considering the solutions to~\eqref{eq:EoMSchrodinger} 
in the limit $r\rightarrow\infty$, which is equivalent to $r^\ast \to \infty$. It this regime, the potential~\eqref{eq:Vpotential} gives a negligible contribution, and the solutions are of the form 
\begin{subequations}
\label{eq:boundarycond}
\be
\label{eq:psirstartoinf}
\lim_{r\to\infty}\psi_{\omega\ell}(r^\ast)= A_{\ell \, {\rm in} }^{\infty} e^{i\omega{r^\ast}}+A_{\ell \, {\rm out} }^{\infty} e^{-i\omega{r^\ast}}. 
\ee
Here, the terms with $A_{\ell \, {\rm in/out} }^{\infty}$ represent an incoming/outgoing wave at infinity, as can be seen by using the radial part~\eqref{eq:psirstartoinf} in the full solution~\eqref{eq:phiSchwans}.Recall that, although we have not included a subindex $\omega$ for simplicity, $A_{\ell \, {\rm in/out} }^{\infty}$ still has a dependence on the frequency.

Near the horizon, $r\rightarrow{r_H}$ or equivalently $r^\ast \to -\infty $ implies from~\eqref{eq:fdef} that $f\to 0$. Since the potential~\eqref{eq:Vpotential} is proportional to $f$ it also vanishes. Thus, the general solutions in the near-horizon limit are also waves, however, due to the nature of the horizon, there can be no outgoing solutions. The boundary condition at the horizon is that the outgoing components vanish and only purely incoming waves remain
\be
\lim_{r\to r_H}
\psi_{\omega \ell}(r^\ast)=A_{\ell \, {\rm in} }^{H} e^{i\omega{r^\ast}} .
\ee
\end{subequations}

We will use these boundary conditions in determining explicit solutions in these two asymptotic regimes, starting with the near-horizon limit, and working in the approximation $M\omega\ll1$. In this section, $M$ denotes the mass of the black hole. The near-horizon region is then characterized by ${r}-{r_H}\ll 1/\omega$, while far from the black hole  $r-{r_H}\gg{M}$. Once we compute our solutions in these regimes we will be able to perform a matched asymptotic expansion where these two regimes overlap.

\subsubsection{Near-horizon solution}

As we will be interested in matching the near-horizon information with the asymptotics at large distances from the black hole, it is convenient to work with the equation of motion in the usual Schwarzschild coordinates from~\eqref{eq:Scalaractionnormalcoords}. 
It is also useful to perform a rescaling of the field 
\be \label{eq:Rofrdef}
\psi_{\omega\ell}(r)=\alpha R_{\omega\ell}(r).
\ee
We substitute~\eqref{eq:Rofrdef} into~\eqref{eq:Scalaractionnormalcoords} and specialize to the limit ${r}-{r_H}\ll 1/\omega$. This leads to the equation of motion
\begin{align}
\label{eq:rmess1}
f r^{\dhat+1}&\p_r\left(f r^{\dhat+1}R_{\omega\ell}'(r)\right)\nonumber\\&-\left(r^{2\dhat}f\dhat^2\lhat(\lhat+1)-\omega^2r_H^{2\dhat+2}\right)R_{\omega\ell}(r)=0~,
\end{align}
where we have used that close to the horizon $\omega{r}\sim\omega{r_H}$.

To cast the differential equation in a solvable form we change coordinates to using $f$ defined in~\eqref{eq:fdef} as the dependent variable. 
Applying this change of variable to~\eqref{eq:rmess1} leads to
\begin{align}
f(1-f)&R_{\omega\ell}''(f)+(1-f)R_{\omega\ell}'(f)\nonumber\\&-\left[\frac{\lhat(\lhat+1)}{(1-f)}-\frac{r_H^2\omega^2}{\dhat^2}\left(\frac{1-f}{f}\right)\right]R_{\omega\ell}(f)=0~.
\end{align}
This differential equation can be transformed into a hypergeometric differential equation by expressing the field as 
\be
\label{eq:RofZ}
R_{\omega\ell}(f)=f^{i\frac{\omega{r_H}}{\dhat}}(1-f)^{\lhat+1}G_{\omega \ell}(f),
\ee
which leads to
\begin{align}
\label{eq:hyper1}
 0=&f (f-1)G_{\omega \ell}''(f)-\left[c^+-f (2 b_\ell+c^+)\right]G_{\omega \ell}'(f) \nonumber\\&+b_\ell \, a^+_\ell\, G_{\omega \ell}(f),
\end{align}
with  
\begin{align}
\label{eq:abcdef}
a^\pm_\ell &=\lhat+1\pm \frac{2ir_H\omega}{\dhat}~,\quad b_\ell=\lhat+1\nonumber\\
c^\pm &=1\pm\frac{2ir_H\omega}{\dhat}.
\end{align}
The solution $G_{\omega \ell}(f)$ to~\eqref{eq:hyper1} is a combination of hypergeometric functions ${}_2F_1(a^+_\ell,b_\ell;c^+;f)$, 
where we follow the conventions of~\cite{AbraStegun}.
In general the second-order differential equation~\eqref{eq:hyper1} has two linearly independent solutions, and the general solution is a linear combination of them. However, special cases of the coefficients~\eqref{eq:abcdef} lead to degeneracies between the two solutions. Specifically, the degeneracy occurs when any of the coefficients $a^+_\ell,\, b_\ell$ or the differences $(c^+-a^+_\ell), \, (c^+-b_\ell)$ are integers. As the frequency $\omega$ is generic, we see from~\eqref{eq:abcdef} that degeneracies arise from integer values for $b_\ell$ when $\lhat$ is a half-integer, and also from $(c^+-a^+_\ell)$ when $\lhat$ is an integer.  
We will start with the case $\lhat$ integer and then distinguish two different analytic continuations of the same solution for $\lhat$ half- and non-integer.

For $\hat \ell \in \mathbb{Z}$ the degenerate solution is given by~\cite{bateman1953higher}\cite{AbraStegun}
\begin{align}
\label{eq:nearhorizoninteger}
G_{\omega \ell}(f)&=(1-f)^{2\lhat+1}{}_2F_1\left(-\lhat,1-a^-_\ell;c^+,f\right)\nonumber\\&=(1-f)^{2\lhat+1}\sum_{n=0}^{\lhat}\frac{(\lhat)_n(1-a^-_\ell)_n}{(c^+)_n}\frac{f^n}{n!}~,
\end{align}
where $c^+$ was defined in~\eqref{eq:abcdef} and
\be
(y)_n=\frac{\Gamma(y+n)}{\Gamma(y)}.
\ee 
denotes the Pochhammer symbol~\cite{Arfken}.

When $\lhat$ is not an integer, the solution is given by~\cite{AbraStegun}
\bea
G_{\omega \ell}(f)&=&e^{2\pi\frac{\omega r_H }{\dhat}}f^{-i\frac{\omega r_H}{\dhat}}A_{\ell\; \rm out}^{H}~{}_2F_1\left(a^-_\ell,b_\ell;c^-_\omega;f\right)\nonumber\\
&&+A_{\ell \; \rm in}^H~{}_2F_1\left(a^+_\ell,b_\ell;c^+;f\right), \quad \lhat \not\in \mathbb{Z} ,
\eea
where $a^\pm_\ell, b_\ell, c^\pm$ were defined in~\eqref{eq:abcdef}.

Using the horizon boundary condition of no outgoing waves, which implies $A_{\ell\; \rm out}^H=0$, we obtain for the  full radial function~\eqref{eq:RofZ}  
\begin{align}\label{eq:fullsolutionhalfintnonint}
&R_{\omega\ell}(r)=
{A_{\ell\; \rm in}^H}~f^{i\frac{r_H\omega}{\dhat}}(1-f)^{\lhat+1}\,{}_2F_1\left(a^+_\ell,b_\ell;c^+;f\right), \;\; \lhat \not\in \mathbb{Z} .
\end{align}

\subsubsection{Asymptotic wave solutions at distances much larger than the black hole's size}
Having obtained the results for the behavior of the near-horizon solutions for scalar perturbations of a black hole, we proceed by establishing its link to the asymptotic wave solutions obtained in the regime $r_H/r \ll 1$. We introduce the parameter 

\be
\label{eq:epsilondef}
\epsilon\equiv \frac{r_H}{r} ,
\ee
and analyze the equation of motion~\eqref{eq:EoMSchrodinger} to first order in $\epsilon$. We choose to work with $r^\ast$ since the equation of motion reduces to a Schr{\"o}dinger-like equation~\eqref{eq:EoMSchrodinger},
which in the limit $r^\ast\rightarrow{\infty}$ reduces to a wave equation
with solution~\eqref{eq:psirstartoinf}.

To connect with the near-horizon solution requires solving for the relation between $r$ and $r^\ast$. We choose to work perturbatively in $\epsilon^{\dhat}$ instead of $\epsilon$ since it is otherwise not possible to expand $1/f$ in the definition~\eqref{eq:rstardef}. Working perturbatively to linear order in $\epsilon^{\dhat}$ we obtain 
\begin{align}
\label{eq:rofrstarasymptotic}
r=r^\ast\left(1+\frac{{\epsilon^\ast}^{\dhat}}{\dhat-1}+\mathcal{O}[({{\epsilon^\ast}^{\dhat}})^2]\right)~,
\end{align}
where we have defined 
\be
\epsilon^\ast=\epsilon\mid_{r=r^\ast},
\ee
with $\epsilon$ given by~\eqref{eq:epsilondef}.

It is interesting to note the simplicity of the result in~\eqref{eq:rofrstarasymptotic} for arbitrary dimensions. This is in contrast with the result for $\dhat=1$, where a logarithm  appears in the relation between $r$ and $r^\ast$ in Schwarzschild spacetime:
\begin{align}
\label{eq:rofrstarasymptoticd1}
r=r^\ast\left(1-\epsilon^\ast\log(r^\ast)+\mathcal{O}({{\epsilon^\ast}}^2)\right)~,
\end{align}
where one has to apply L'H{\^o}pital's rule to~\eqref{eq:rofrstarasymptotic} and take the limit $\dhat\rightarrow1$ together with the small-size limit $r_H\rightarrow0$,
\begin{align}
\lim_{r_H\rightarrow0}\lim_{\dhat\rightarrow1}\frac{\frac{d}{d\dhat}{\epsilon^\ast}^{\dhat}}{\frac{d}{d\dhat}(\dhat-1)}
&=\lim_{r_H\rightarrow0}\left(\epsilon^\ast\log(r_H)-\epsilon^\ast\log(r^\ast)\right)\nonumber\\&=-\epsilon^\ast\log(r^\ast)~.
\end{align}
Here $\epsilon$ is defined in~\eqref{eq:epsilondef} and we have kept only the leading order term.

Altogether, we find that the limit of the radial solution for generic $\dhat$ and $\lhat$ vanishes. This means that the $\epsilon^\ast$ corrections do not introduce any divergence and therefore we can safely use the flat space solution with $\epsilon^\ast=0$.
This also confirms the flat-space approximation used in the effective theory side when $\dhat$ and $\lhat$ are generic complex numbers.
Using that in this limit $\epsilon^\ast=0$, the asymptotic wave solution for distances much larger than the size of the black hole will be given by~\eqref{eq:EoMSchrodinger} with $f=1$ and $r^\ast=r$,
\begin{align}\label{eq:RadialSolAsympt}
R_{\omega\ell}(r)=&r^{-\dhat/2}\left({A}_{\ell \; \rm reg}^{  \infty}J_{\dhat/2+\ell}(\omega{r})+A_{\ell \; \rm irreg}^{ \infty} Y_{\dhat/2+\ell}(\omega{r})\right),
\end{align}
where we have chosen the regular/irregular basis rather than the in/out states. If we now look at the boundary condition at infinity \eqref{eq:psirstartoinf}, we see that, given $r=r^\ast$, is the same as in \eqref{eq:cinoutdef}.

\subsubsection{Determining the imprint of the black hole's properties in the scattering amplitudes}

To complete the calculation of the response function we next compute the ratio $A^{\infty}_{\ell \; \rm in}/A^{ \infty}_{\ell \; \rm out}$ in terms of properties of the perturbed black hole using matched asymptotic expansions. 
 Specifically, we will consider the asymptotic expansion of the near-horizon solution~\eqref{eq:fullsolutionhalfintnonint} for large $1/\epsilon$ and of the asymptotic solution~\eqref{eq:RadialSolAsympt} for $\omega r\ll 1$. The near-horizon region is ${r}-{r_H}\ll 1/\omega$, while the far-zone region is $r-{r_H}\gg{M}$. The matching is performed where the two asymptotic expansions overlap, and with the use of analytic continuation in $\lhat$; see Fig.~\ref{fig:Calculation} for an illustration of the process.

We note that only the in- and outgoing solutions are well-defined physical states. However, as in Sec.~\ref{sec:EFT}, it is easier to compute the ratio of the wave amplitudes in the regular/irregular basis, with $A^{ \infty}_{\omega\ell \; \rm irreg}/A^{ \infty}_{\omega\ell \; \rm reg}$ understood as constants defined by~\eqref{eq:Cregiregtoinout}.

We first consider the asymptotic expansion of the solution~\eqref{eq:RadialSolAsympt} for $\omega(r-r_H)=\omega{r}(1-\epsilon)\sim\omega{r}\ll1$. The limiting behavior of the Bessel functions is given by~\cite{NIST:DLMF}
\begin{align}
\lim_{z\ll 1}J_{\nu}(z)&\rightarrow\left(\frac{z}{2}\right)^\nu\frac{1}{\Gamma[\nu+1]}~,\\
\lim_{z\ll 1}Y_{\nu}(z)&\rightarrow-\left(\frac{z}{2}\right)^{-\nu}\frac{\Gamma[\nu]}{\pi}.
\end{align}
The radial solution~\eqref{eq:RadialSolAsympt} thus becomes
\begin{align}
\lim_{\omega r\ll 1}R_{\omega\ell}(r)=&-A^{ \infty}_{\omega\ell \; \rm irreg}\left(\frac{\omega}{2}\right)^{-\frac{\dhat}{2}(2\lhat+1)}\frac{\Gamma(p)}{\pi}r^{-\dhat(\lhat+1)}\nonumber\\&+A^{ \infty}_{\omega\ell \; \rm reg}\left(\frac{\omega}{2}\right)^{\frac{\dhat}{2}(2\lhat+1)}\frac{1}{\Gamma(p+1)}r^{\dhat\lhat},
\end{align}
with 
\be
\label{eq:pdef}
p=\frac{\dhat}{2}(2\lhat+1).
\ee

Next, we consider the asymptotic expansion of the near-horizon solutions in the limit $\epsilon \to 0$, 
with $f=1-\epsilon^{\dhat}$. 
The degenerate solution~\eqref{eq:nearhorizoninteger} for integer arguments behaves as~\cite{AbraStegun}
\begin{align}
\lim_{\epsilon\to 0}R_{\omega\ell}(r)\propto \left(\frac{1}{\epsilon}\right)^{\dhat\lhat}, \quad \lhat \in \mathbb{Z}
\end{align}
This contains only positive powers of $r$ corresponding to growing, regular solutions; a decaying, irregular component is absent. Thus, we conclude that $A^{H}_{\ell \; \rm irreg}=0$ in the limit $\epsilon \to 0$. 
 
The solution for non-integer $\lhat$ is given by~\eqref{eq:fullsolutionhalfintnonint}. Since we have to take the limit $\epsilon\rightarrow0$, or equivalently $f\rightarrow1$, it is useful to use hypergeometric linear transformations in order to change the argument of the hypergeometric function from $f$ to $1-f$. This is useful given that ${}_2F_1(a,b;c,0)=1$. Since none of the parameters $a,b,c$ are integers, the linear transformation reads~\cite{bateman1953higher}
\begin{align}\label{eq:lineartransfnonint}
{}_2&F_1(a,b;c;x)=(1-x)^{-a-b+c} \frac{\Gamma(c) \Gamma (a+b-c)}{\Gamma\left(a\right) \Gamma (b)}  \nonumber\\&\times{}_2F_1(c-a,c-b;-a-b+c+1;1-x)\nonumber\\&+\frac{\Gamma (c) \Gamma (c-a-b)}{\Gamma(c-a)\Gamma (c-b)}{}_2F_1(a,b;a+b-c+1;1-x)~.
\end{align}
Substituting the linear transformation into~\eqref{eq:fullsolutionhalfintnonint} and taking the limit $f\rightarrow1$ with $(1-f)=\epsilon$ fixed yields
\begin{align}\label{eq:NearHSolnonint}
\lim_{\epsilon\to 0}&R_{\omega\ell}(r)=A_{\ell \, {\rm in}}^H\frac{\Gamma(-2\lhat-1)\Gamma(c^+)}{\Gamma(-\lhat)\Gamma(1-a^-_\ell)}\epsilon^{\dhat(\lhat+1)}\nonumber\\&+A_{\ell \, {\rm in}}^H\frac{\Gamma(2\lhat+1)\Gamma(c^+)}{\Gamma(b_\ell)\Gamma(a^+_\ell)}\left(\frac{1}{\epsilon}\right)^{\dhat\lhat}, \;\; \lhat \not\in \mathbb{Z}, \mathbb{Z}/2
\end{align}
 
We next consider the case where $\lhat$ is half-integer. The solution for this case is also given by~\eqref{eq:fullsolutionhalfintnonint}. For the case $\lhat\in\mathcal{Z}/2$, $c-a-b=-m=-2\lhat-1$ is a negative integer and the linear transformation~\eqref{eq:lineartransfnonint} develops a pole. The linear transformation is then computed by analytic continuation and is given by~\cite{AbraStegun}\cite{bateman1953higher}
\begin{align}\label{eq:linartransfhalfint}
{}_2&F_1(a,b;a+b-m;x)=(1-x)^{-m} \frac{\Gamma(m) \Gamma (a+b-m)}{\Gamma\left(a\right) \Gamma (b)}  \nonumber\\&\times\sum_{n=0}^{m-1}\frac{(b-m)_n(a-m)_n}{(1-m)_nn!}(1-x)^{n}\nonumber\\&+(-1)^m\frac{\Gamma (a+b-m)}{\Gamma(a-m)\Gamma (b-m)}\sum_{n=0}^{\infty}\frac{(a)_n(b)_n}{(n+m)!n!}\nonumber\\&\quad\times\left[\kappa''_n-\log(1-x)\right](1-x)^{n}~.
\end{align}
where 
\be
\label{eq:kappadef}
\kappa''_n=\psi(1+m+n)+\psi(1+n)-\psi(a+n)-\psi(b+n)
\ee
and 
\be
\psi(x)=\frac{\Gamma'(x)}{\Gamma(x)}\ee is the digamma function. Substituting into~\eqref{eq:fullsolutionhalfintnonint} yields
\begin{align}
\label{eq:Rinfnonintfull}
&\frac{R_{\omega\ell}(r)}{A_{\ell \, {\rm in} }^{ H}f^{i\omega r_H/\dhat}}
=\frac{\Gamma(2\lhat+1)\Gamma(c^+)}{\Gamma(a^+_\ell)\Gamma(b_\ell)}
\nonumber\\ &\quad\times\sum_{n=0}^{2\lhat}\frac{(-\lhat)_n(1-a^-_\ell)_n}{(-2\lhat)_nn!}(1-f)^{n-\lhat}\nonumber\\&\; \;+\frac{(-1)^{2\lhat+1}\Gamma\left(c^+\right)}{\Gamma(1-a^-_\ell)\Gamma(-\lhat)}\sum_{n=0}^{\infty}\frac{(a^+_\ell)_n(b_\ell)_n}{(n+2\lhat+1)!n!}\nonumber\\&\quad\times\left[\kappa''_n-\log(1-f)\right](1-f)^{n+\lhat+1}, \quad \lhat \in \mathbb{Z}/2 .
\end{align}
 Extracting the dominant powers of $\epsilon$ and $1/\epsilon$ in the two series in~\eqref{eq:Rinfnonintfull} we obtain
\begin{align}\label{eq:NearHSolhalfint}
&\lim_{\epsilon \to 0}R_{\omega\ell}(r)=A_{\ell \, {\rm in}}^H f^{i\frac{r_H\omega}{\dhat}}\frac{\Gamma (2\lhat+1)\Gamma(c^+)}{\Gamma(a^+_\ell)\Gamma(b_\ell)}\left(\frac{1}{\epsilon}\right)^{\dhat\lhat}+\ldots\nonumber\\&\quad+A_{\ell \, {\rm in}}^Hf^{i\frac{r_H\omega}{\dhat}}\frac{(-1)^{2\lhat+1}\Gamma(c^+)}{\Gamma(1-a^-_\ell)\Gamma(-\lhat)(2\lhat+1)!}\nonumber\\&\quad\times\left[\kappa''_0-\dhat\log\left(\epsilon\right)\right]\epsilon^{\dhat(\lhat+1)}+\dots~, \quad \lhat \in \mathbb{Z}/2 ,
\end{align}
where the dots denote higher positive or negative powers of $\epsilon$. The appearance of the logarithm is in agreement with \cite{Kol:2011vg} and \cite{Hui:2020xxx}, where they argue that it is a consequence of a classical renormalization group flow of general relativity. However, as discussed below, we will work with generic, real values of $\lhat$ and only in the end take the limit $\lhat\rightarrow\mathbb{Z}/2$. When taking this limit we obtain a logarithm of $r_H$, and no $r$-dependent coefficients.

As we will be interested only in a matching to the leading order, where the two asymptotics we are considering manifestly exhibit an overlap, it will not be necessary to introduce scaled matching coordinates. Hence, we can substitute back our definition of $\epsilon$ given in~\eqref{eq:epsilondef} into the near-horizon solutions~\eqref{eq:NearHSolnonint} and~\eqref{eq:NearHSolhalfint}. Next, we can directly perform the matching of the near-horizon solution and the asymptotic solution by considering the coefficients in front of each radial dependence, i.e. $r^{-\dhat(\lhat+1)}$ and $r^{\dhat\lhat}$. For generic $\lhat\in\mathbb{R}$, not an integer or half-integer, this matching yields
\be
\label{eq:Ainftyratio}
\frac{A^\infty_{\ell \, {\rm irreg}}}{A^\infty_{\ell \, {\rm reg}}}=-\frac{\pi \left(\omega r_H/2\right)^{\dhat(2\lhat+1)}\Gamma(-2\lhat-1)\Gamma(b_\ell)\Gamma(a^+_\ell)}{\Gamma(-\lhat)\Gamma(2\lhat+1)\Gamma(p)\Gamma(p+1)\Gamma(1-a^-_\ell)}, 
\ee
with 
$p$ given in~\eqref{eq:pdef}, $a^\pm_\ell$ and $b_\ell$ defined in~\eqref{eq:abcdef}, and $\epsilon$ given in~\eqref{eq:epsilondef}. Note that when specializing to integer $\lhat$ at the level of the matching, one obtains that~\eqref{eq:Ainftyratio} is zero. Half-integer $\lhat$ leads to a different functional form of this ratio, similar to the static case discussed in~\cite{Kol:2011vg,Hui:2020xxx}.
This arises because for integer $\lhat$, the hypergeometric function characterizing the near-horizon solution becomes the degenerate solution~\eqref{eq:nearhorizoninteger}, while for half-integers it develops poles \eqref{eq:NearHSolhalfint}. The problems with considering these singular cases directly are avoided by using analytic continuation in $\lhat$. Keeping $\lhat$ generic enables us to work with the finite, well-behaved result~\eqref{eq:Ainftyratio}, and the singular cases 
are obtained by carefully taking the limits $\lhat\rightarrow\mathbb{Z}$ and $\lhat\rightarrow\mathbb{Z}/2$ of the final, generic results.

Note that in the case of static tides, one essentially only has the near-horizon part of the solution, making it more difficult to extract gauge-invariant information asymptotically.
Here instead we can make the connection to gauge-invariant scattering data based on in- and out-going wave solutions.
This is also a more physical setup, since even for adiabatic tides the frequency of the tidal field is never exactly zero in an astrophysical environment.

\subsection{Matching to the skeletonized effective action description}
In this section, we address the final step in obtaining the tidal response function of a black hole by connecting the information about the perturbed black hole contained in the scalar-wave amplitudes as computed in Sec.~\ref{sec:Schwarzschild} with the definition of the response function from Sec.~\ref{sec:EFT}. This requires an identification between the asymptotic waves in Schwarzschild and Minkowski spacetimes. To facilitate this link in a coordinate-invariant manner, we will base the identification on the geometry of light cones, as discussed below.

\subsubsection{Identification of the null infinities of Schwarzschild and Minkowski spacetimes }

To connect with the effective action from Sec.~\ref{sec:EFT} requires the limit of the perturbative calculations from Sec.~\ref{sec:Schwarzschild} when the black hole is viewed from distances much larger than its size and shrinks to nearly a point, $r_H\rightarrow0$. When taking this limit we recover an asymptotically nearly flat spacetime. In the effective action discussed in Sec.~\ref{sec:EFT}, we assumed a Minkowksi spacetime for simplicity. In general there would be corrections to the metric potentials in powers of $1/r$. In principle, these should be included and must match to the Schwarzschild asymptotics near null infinity. Here, we only capture the leading-order behavior in this regime, which will be sufficient for our purposes. 

To make the asymptotic identification between the Schwarzschild and Minkowski spacetimes we use  double-null coordinates $u,v$. 
For Schwarzschild spacetime, they are defined by
\begin{subequations}\label{eq:doublenullcoord}
\begin{align}
du&=dt-\frac{1}{f}dr=dt-dr^\ast,\\
dv&=dt+\frac{1}{f}dr=dt+dr^\ast.
\end{align}
\end{subequations}
In Minkowski spacetime, these coordinates reduce to $u=t-r$ and $v=t+r$. 
Such coordinates are adapted to radial null geodesics and therefore along the light cones. Since light cones have an intrinsic geometric meaning and are invariant objects asymptotically, this set of coordinates enables a robust identification between the incoming and outgoing solutions both in the effective theory and the black hole perturbation calculations.

We first discuss the solutions in the effective theory expressed in null coordinates. The asymptotic solutions for in- and outgoing waves were obtained in~\eqref{eq:phiinoutEFTasymptotic}. The characteristic are exactly along $u,v$, and thus, the dependence on $(t\pm r)$ can immediately be transformed to the null coordinates using their flat-space definition. This yields
\begin{align}
\label{eq:EFTofuvasy}
\lim_{r\rightarrow\infty}\phi(u,v)=
\frac{C^L_{\rm in} n_L( i\omega)^\ell e^{i\omega v}}{r(u,v)^{\frac{\dhat+1}{2}}}+
\frac{C^L_{\rm out} n_L(- i\omega)^\ell e^{i\omega u}}{r(u,v)^{\frac{\dhat+1}{2}}},
\end{align}
with $r(u,v)=(v-u)/2$ in flat space.

\begin{figure}[t]
	\begin{center}			\includegraphics[width=0.3\textwidth,clip]{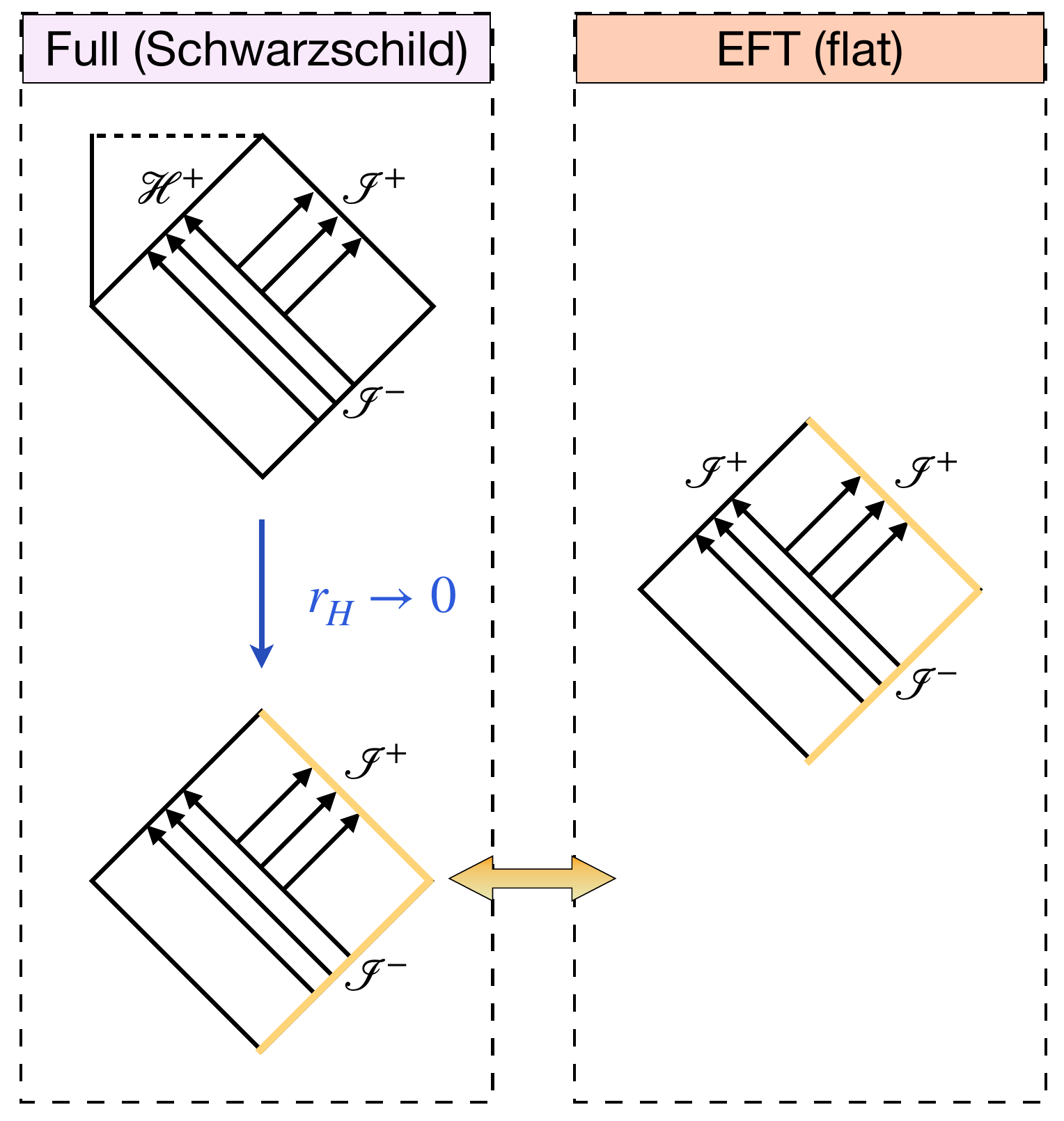}
		\caption{\label{fig:Penrosediagram}
		\emph{Penrose diagrams illustrating the asymptotic identification of null infinities} $(\mathscr{I}^\pm)$ of the Schwarzschild spacetime in the limit that the size of the black hole $r_H$ shrinks to zero and the Minkowski spacetime we use in the effective description. The matching is performed using double null coordinates adapted to the light cones.}
	\end{center}
\end{figure}

For the Schwarzschild case, it is easiest to consider the asymptotic form of the equations of motion instead of transforming the solutions. Instead of the previous ansatz~\eqref{eq:phiSchwans}, we now decompose the scalar field as
\begin{align}
\phi=\sum_{\ell m}{\chi}_{\ell m}(u,v)Y_{\ell m}(\Omega). 
\end{align}
Substituting this ansatz into the action~\eqref{eq:actionscalar} and using the metric~\eqref{eq:Schw} transformed to null coordinates through~\eqref{eq:doublenullcoord} leads to the following equation of motion for ${\chi}(u,v)$
\begin{multline}
\label{eq:uveom}
2r^{\dhat+1}\p_u\p_v\chi+\p_v\left(r^{\dhat+1}\right)\p_u\chi+\p_u\left(r^{\dhat+1}\right)\p_v\chi\\+2fr^{\dhat-1}\dhat^2\lhat(\lhat+1)\chi=0,
\end{multline}
where $r=r(u,v)$ is defined through~\eqref{eq:doublenullcoord} and $u=t-r^\ast$, $v=t+r^\ast$. In the limit $\epsilon\to 0$ where the black hole shrinks to a point or equivalently $r\to \infty$, the factor $f$ in the second line of~\eqref{eq:uveom} becomes unity ($f\to 1$). For large $r$, the most dominant term in the differential equation~\eqref{eq:uveom} is the first one, and has a form identical to the flat-space wave equation in null coordinates. 
Thus, we can write down the solution to~\eqref{eq:uveom} in terms of in- and outgoing spherical waves in the asymptotic regime $\epsilon\to 0$ as
\be
\label{eq:phiSchwuv}
\lim_{\epsilon\to 0}\phi\sim \sum_{\ell m} \left[A^{\infty}_{\ell\, {\rm in}}\frac{e^{i\omega{v}}}{{r(u,v)}^{\frac{\dhat+1}{2}}}+A^{\infty}_{\ell\, {\rm out}}\frac{e^{i\omega{u}}}{{r(u,v)}^{\frac{\dhat+1}{2}}}\right]Y_{\ell m}(\Omega).
\ee
As discussed in Sec.~\ref{sec:Schwarzschild}, for $\epsilon\to 0$ we also have that $r^\ast\rightarrow{r}$. This implies that the light cones and correspondingly the $u,v$ coordinates of the Schwarzschild and Minkowski spacetime coincide asymptotically for $\epsilon=0$. To this first approximation we are considering, we can identify both the future and past null infinities between the effective and Schwarzschild descriptions, and use this to relate the results of the two different  calculations. Figure~\ref{fig:Penrosediagram} illustrates this reasoning. As mentioned above, in general, higher-order corrections would be included in this identification. 
Thus, using the conversion of the coefficients from the STF to the spherical harmonic basis from~\eqref{eq:inouttospher} in~\eqref{eq:EFTofuvasy}, applying the identity~\eqref{eq:Yton}, and comparing with~\eqref{eq:phiSchwuv} leads to the trivial identification
\begin{align}\label{eq:CEFTFullmatching}
C_{\ell \, {\rm in/out} }=A^\infty_{\ell \, {\rm in/out}}~,
\end{align}
where $C_{\ell \, {\rm in/out} }$ are the coefficients of the effective field theory solution and $A^\infty_{\ell \, {\rm in/out} }$ are the coefficients of the asymptotic solution in the Schwarzschild spacetime.

With this identification, we can compute the coefficients in the full theory via analytical methods such as matched asymptotic expansion (facilitated by analytic continuation), or via numerical methods. We note that the amplitudes $A^\infty_{\ell \, {\rm in/out}}$ need not be obtained in generic dimension, which may indeed be computationally unfeasible for rotating compact objects and/or numerical approaches. However, when matching the asymptotic waves in four spacetime dimensions, the background spacetime curvature must be taken into account, as indicated by the logarithm in~\eqref{eq:rofrstarasymptoticd1} which introduces an infrared-singular contribution to the phase. The double-null coordinates streamline the matching by absorbing such contributions, making the agreement of the infrared/asymptotic physics between full and effective theory manifest.

\subsubsection{Explicit results for the response function}
With the above results, we can compute the explicit expression for the frequency-dependent response function of the black hole to scalar tidal perturbations. We first obtain the response for generic $\lhat$, and verify that in the static limit $\omega\to 0$ this agrees with previous results~\cite{Kol:2011vg,Hui:2020xxx}. We then discuss the special singular cases when $\lhat$ is integer, which is relevant for four spacetime dimensions, and half-integer by carefully taking the limits of the general result. 

The response is obtained by substituting~\eqref{eq:Ainftyratio} into~\eqref{eq:responsefunction}, which yields
\be
\label{eq:responsenonsimpl}
F_{\ell}(\omega)=\frac{2^{2-\ell}\pi^{\dhat/2+1}r_H^{\dhat(2\lhat+1)}\Gamma(b_{\ell})\Gamma(-2\lhat-1)\Gamma(a^+_{\ell})}{\Gamma(-\lhat)\Gamma(2\lhat+1)\Gamma(p)\Gamma(1-a^-_{\ell})}~,
\ee
where we use $\ell=\lhat\dhat$, and the parameters $a_\ell^\pm$ and $p$ are defined in \eqref{eq:abcdef} and \eqref{eq:pdef}. This expression can be simplified by using the Legendre multiplication formula for the gamma functions \cite{NIST:DLMF},
\be\label{eq:Legendremult}
\Gamma(2z)=\pi^{-1/2}2^{2z-1}\Gamma(z)\Gamma(z+\frac{1}{2})~,
\ee 
which leads to
\be\label{eq:responsesimpl}
F_{\lhat}(\omega)=\frac{\pi^{\frac{\dhat}{2}+1}\Gamma(-\lhat-\frac{1}{2})\Gamma(a^+_\ell)r_H^{\dhat(2\lhat+1)}}{2^{(4+\dhat)\lhat}\Gamma(\lhat+\frac{1}{2})\Gamma(p)\Gamma(1-a^-_\ell)}~.
\ee
We now consider the limit $\omega\rightarrow0$ of \eqref{eq:responsesimpl}, which yields the static Love numbers. Using the reflection formula \cite{NIST:DLMF}
\bea
\Gamma(z)\Gamma(1-z)=\frac{\pi}{\sin(\pi z)}~,
\eea 
it holds
\begin{align}\label{eq:Gammatotan}
\frac{\Gamma\left(-\lhat-\frac{1}{2}\right)}{\Gamma(-\lhat)}=\frac{\Gamma(\lhat+1)}{\Gamma\left(\lhat+\frac{3}{2}\right)}\tan(\pi\lhat)~.
\end{align}
Substituting \eqref{eq:Gammatotan} into  \eqref{eq:responsesimpl} and using \eqref{eq:pdef} yields
\begin{align}\label{eq:responseKS}
&F_{\lhat}(\omega)=\frac{\pi^{\frac{\dhat}{2}+1}\Gamma^2(\ell+1)r_H^{\dhat(2\lhat+1)}\tan(\pi\lhat)}{2^{(4+\dhat)\lhat}\Gamma\left(\frac{\dhat}{2}(2\lhat+1)\right)\Gamma(\lhat+\frac{1}{2})\Gamma(\lhat+\frac{3}{2})}~,
\end{align}
in agreement with~\cite{Kol:2011vg}.
As discussed in~\cite{Kol:2011vg}, this expression has poles for specific values of $\lhat$, which play the role of counterterms in the effective action. That is, though we based the matching on a calculation in the effective theory in a flat background, we can extract information about poles appearing at higher orders (curved background) in the effective theory through analytic continuation, which is an impressive display of its power.

Having confirmed that the static limit of the response~\eqref{eq:responsesimpl} reproduces previous results, we next examine the full frequency-dependence in the limit $\lhat\rightarrow\mathbb{Z}$ relevant for four dimensions. Using the definitions of $a_\ell^\pm$ from~\eqref{eq:abcdef} and the identities
\begin{align}
\Gamma&\left(a_\ell^+\right)=\Gamma\left(1+\frac{2ir_H\omega}{\dhat}\right)\prod_{k=1}^{\lhat}\left(k+\frac{2i{r_H}\omega}{\dhat}\right)~,\\
\Gamma&\left(1-a_\ell^-\right)=\frac{\Gamma\left(1+\frac{2ir_H\omega}{\dhat}\right)}{\frac{2ir_H\omega}{\dhat}(-1)^{\lhat}\prod_{k=1}^{\lhat}\left(k-\frac{2i{r_H}\omega}{\dhat}\right)}
\end{align}
in~\eqref{eq:responsesimpl} leads to
\begin{align}\label{eq:responseprod}
F_{\lhat\in\mathbb{Z}}(\omega)&=i\omega\frac{(-1)^{\lhat}2\pi^{\frac{\dhat}{2}+1}\Gamma\left(-\lhat-\frac{1}{2}\right)r_H^{\dhat(2\lhat+1)+1}}{ 2^{(\dhat+4)\lhat}\dhat\, \Gamma\left(\lhat+\frac{1}{2}\right)\Gamma\left(p\right)}\nonumber\\
&\times \prod_{k=1}^{\lhat}\left(k^2+\frac{4r_H^2\omega^2}{\dhat^2}\right).
\end{align}

Finally, we consider the special case that $\lhat\rightarrow\mathbb{Z}/2$. In this case the response function~\eqref{eq:responsenonsimpl} diverges due to the presence of simple poles in $\Gamma(-2\lhat-1)$. We can solve this issue, as done similarly in \cite{Kol:2011vg}, by expanding in $2\lhat=n-\varepsilon$ with $n$ an integer and $\varepsilon\rightarrow0$ a small parameter, which isolates the finite contribution. We use the property of the $\Gamma$ function
\cite{Arfken}
\begin{align}
\Gamma(-k+\varepsilon)=\frac{(-1)^{k}}{k!\varepsilon}+\mathcal{O}(\varepsilon^0)~
\end{align}
for any integer $k$.
In the response~\eqref{eq:responsesimpl}, the factor $\Gamma(-2\lhat-1)$ appears together with $r_H^{\dhat(2\lhat+1)}=r_H^{2\lhat\dhat}r_H^{\dhat}$. The first of these can be written as
 $r_H^{2\lhat\dhat}=r_H^{(n-\varepsilon)\dhat}=r_H^{n\dhat}r_H^{-\varepsilon\dhat}$. The last factor here must be included when considering the limit $\varepsilon\to 0$ of the divergences in the response. Introducing the cutoff scale $\Lambda$, and defining $\hat{r}_H=r_H/\Lambda$, the expansion of the relevant pieces of the response in this limit truncated at $O(\varepsilon^0)$ is then given by
\begin{align}
\Gamma&(-n-1+\varepsilon)\hat{r}_H^{-\dhat\varepsilon}
=\frac{(-1)^{2\lhat+1}}{(2\lhat+1)!}\frac{1}{\varepsilon}\left[1-\varepsilon\dhat\log\left(\hat{r}_H\right)\right]\nonumber\\&=-\dhat\log\left(\hat{r}_H\right)\frac{(-1)^{2\lhat+1}}{(2\lhat+1)!}+\frac{(-1)^{2\lhat+1}}{(2\lhat+1)!}\frac{1}{\varepsilon}.
\end{align}
Hence, only the first term is finite in the limit $\varepsilon\to 0$ and should be considered to describe the response function, while the divergent part should be interpreted as a counterterm in the action~\cite{Kol:2011vg}.\footnote{An explicit systematic construction of the internal action $S_\text{int}$ in terms of modes degrees of freedom as in~\cite{Gupta:2020lnv} might provide a cleaner split between the dynamical mode response and counterterms.} With this convention, the response function for half-integer $\lhat$ reads
\begin{align}
\frac{F_{\lhat\in\mathbb{Z}/2}(\omega)}{\Lambda^{2\dhat(\lhat+1)}}&=\frac{(-1)^{2\lhat}2^{2-\ell}\dhat\, \pi^{\dhat/2+1} \hat{r}_H^{2\dhat(\lhat+1)}\Gamma(b_{\ell})\Gamma(a^+_{\ell})\log(\hat{r}_H)}{\Gamma(2\lhat+2)\Gamma(-\lhat)\Gamma(2\lhat+1)\Gamma(p)\Gamma(1-a^-_{\ell})}\nonumber\\&+\frac{(-1)^{2\lhat+1}2^{2-\ell}\, \pi^{\dhat/2+1} \hat{r}_H^{2\dhat(\lhat+1)}\Gamma(b_{\ell})\Gamma(a^+_{\ell})}{\Gamma(2\lhat+2)\Gamma(-\lhat)\Gamma(2\lhat+1)\Gamma(p)\Gamma(1-a^-_{\ell})}\frac{1}{\varepsilon}.
\end{align}

\subsubsection{Love numbers and absorption encoded in the response}
From \eqref{eq:responseprod} one can see that when $\lhat$ and $\dhat$ are integer numbers, the real part of the response function vanishes at all orders in $\omega$ within the approximation $M\omega\ll1$. Hence, for a four-dimensional nonrotating black hole, not only the static Love number vanishes, but the entire real part of the response function,
\begin{align}
\Re\{F_{\lhat\in\mathbb{Z}}(\omega)\}=0~.
\end{align}

Furthermore, we can also compute the absorption cross section and compare with the result from~\cite{Page:1976df} for $\dhat=1$ and $\lhat=\ell=0$. The definition of the partial absorption cross section is given by
\begin{align}\label{eq:defabsorb}
\sigma_{\rm abs}^{\ell}=\frac{\pi}{\omega^2}\left(2\ell+1\right)\left(1-\left|\frac{A_{\rm out}}{A_{\rm in}}\right|^2_{\ell}\right)~.
\end{align}
Using~\eqref{eq:Cregiregtoinout} for $\ell=0$ we obtain
\begin{align}
\sigma_{\rm abs}^{\ell=0}=16M^2\pi~,
\end{align}
in agreement with the literature. On the other hand, the response function for $\dhat=1$ and $\lhat=0$ reads
\begin{align}
F_{\lhat=0}(\omega)=-4r_H^2i\pi\omega=-16M^2i\pi\omega~.
\end{align}
This result suggests that, in the spirit of~\cite{Goldberger:2005cd} and the optical theorem, the absorption cross section and the imaginary part of the response function are proportional\footnote{We noticed a typo in a previous version of this paper when comparing with the results in \cite{Ivanov:2022qqt}.},
\begin{align}
\left|\Im\{F_{\ell,\dhat=1}(\omega)\}\right|=\frac{(2\ell-1)!!}{\omega^{2\ell-1}}\sigma_{\rm abs}~.
\end{align}
One has to take the absolute value of the imaginary part because the terms with odd powers of the frequency will have a different sign depending on the chosen convention of the Fourier transform.

For generic spacetime dimensions and multipolar order, it holds
\begin{align}
    &\Re\{F_{\lhat}(\omega)\}=0 \iff\Im{\left\{\frac{C_{\ell}^{ \rm in}}{C_{\ell }^{ \rm out}}e^{i\frac{\pi}{2}(\dhat+1)}\right\}}=0\\
    &\Im\{F_{\lhat}(\omega)\}=0 \iff\left|\frac{C_{\ell}^{ \rm in}}{C_{\ell }^{ \rm out}}\right|^2=1~.
\end{align}
We discuss an analogy with optics in Sec.~\ref{sec:discussion}.

\section{Summary and Discussion}
\label{sec:discussion}

An important quantity for gravitational wave signatures of the nature and internal structure of compact objects is its response to tidal perturbations. The response is operationally defined by the imprints on gauge-invariant observables, such as the binding energy as a function of frequency or the ratio of in- and outgoing wave amplitudes at null infinity. These observables are directly computed from an effective action describing the physics at large distances from the object, where it is modeled as a center-of-mass worldline augmented with multipole moments. At that level, the response is defined mathematically as the ratio between 
the induced multipole moments $Q^L$ to the tidal field $E_L$, specifically
\be
Q^L(\omega)=-F_\ell(\omega)E^L(\omega).
\ee
Here, the function $F_\ell(\omega)$ is the complex frequency-dependent response function and all quantities are defined in frequency domain.
In the case of scalar perturbations, the tidal field is given by $E_L= \underset{r\rightarrow{0}}{\text{FP}}\p_L\phi(\omega)$, where $\phi$ is the scalar field. 

Extracting the response function from gauge-invariant observables of a binary system and in particular discriminating its effects from (unknown) higher PN point-mass corrections is subtle yet important to avoid biases in the interpretation. The required distinction can be accomplished in a rigorous way through analytic continuation in the number of spacetime dimensions and/or multipole orders. Consequently, tidal effects can be unambiguously determined without having to carry out high-order PN calculations.

A highly useful framework for computing gauge invariant quantities in a binary system is an effective action description, where the compact objects are reduced to  center-of-mass worldlines with multipole moments. The tidal response imprinted in observables such as the binding energy or gravitational waves is thus directly related to quantities appearing in the effective action, for instance coupling coefficients. Relating the effective action to detailed properties of the compact object requires matching calculations. In particular, one must compute the induced multipoles $Q^L$ defined in the spacetime outside the object for a given microphysical model of its internal structure, and relate the result to the quantities appearing in the effective action. To avoid ambiguities in the matching, it is highly advantageous to establish the link between the perturbative description and the effective action by considering wave scattering states defined at null infinity instead of a stationary setup as in standard approaches.

In scalar wave scattering, the scalar tidal response function is related to the ratio of amplitudes of in- and outgoing waves $C_{\rm in}/C_{\rm out}$ defined at null infinity of Minkowski spacetime by
\be 
F_\ell(\omega)=K_\phi\tilde{F}_\ell(\omega)~.
\ee 
Here $K_\phi$ is the scalar field coupling constant
\be
S_{\phi}=-\frac{K_{\phi}}{2}\int d^dx \sqrt{-g} \; g^{\alpha\beta}~\nabla_\alpha\phi\nabla_\beta\phi,
\ee
related to that of the full theory such that $K_\phi=K_{\phi}^{\rm full}$, and  $\tilde{F}_\ell(\omega)$ the normalised response function
\be
\tilde{F}_\ell(\omega)=i\,\Xi_\ell\left[1-\dfrac{2}{1+\frac{C_{\ell}^{ \rm in}}{C_{\ell }^{ \rm out}}e^{i\frac{\pi}{2}\left(\dhat+1\right)}}\right]
\ee
or, introducing the complex phase shift $\delta_\ell$ defined via $C_{\ell}^{\rm in}/C_{\ell}^{\rm out}= e^{2i\delta_\ell}$,
\be
\label{eq:responsefunctionphase}
\tilde{F}_\ell(\omega)=-\Xi_\ell\tan\left[\delta_\ell+\frac{\pi}{4}(\dhat+1)\right]~,
\ee
with
\be
\label{eq:Xildef2}
\Xi_\ell=-\frac{4\pi^{\dhat/2}}{2^{\ell}}\left(\frac{2}{\omega}\right)^{\dhat+2\ell}\Gamma\left(\frac{\dhat}{2}+\ell+1\right).
\ee
 This result is similar to the frequency-dependent response in optics, with the analog of response being the refractive index of a material.
An imaginary refractive index corresponds to absorption of light and a change in amplitude. By contrast, in the absence of absorption, the refractive index encodes a phase shift of the light and leads to refraction of incident light beams.

The identification of the in- and out scattering states at null infinity of the Minkowski and Schwarzschild spacetimes is made rigorous by basing it on the geometry of the light cones. This reveals that there is a one-to-one correspondence between the in- and outgoing wave amplitudes 
\be
\frac{C_{\rm in}}{C_{\rm out}}\mid_{\rm Minkowski}=\frac{A_{\rm in}^\infty}{A_{\rm out}^\infty}\mid_{\rm Schwarzschild}.
\ee
The above results are valid in general for scalar perturbations to any compact object in GR. The connection to the microphysical properties, however, requires specializing to a particular kind of compact object. 

For a Schwarzschild black hole, it is possible to perform analytical calculations that trace the information flow from the perturbed black hole to regions far from it in the limit $M\omega\ll 1$. Matched asymptotic expansions reveal that the response function is given by 
\be
\label{eq:responsesummary}
F_{\ell}(\omega)=W_{\ell\dhat}\begin{cases}
\dfrac{\Gamma(\lhat+1+2i\omega r_H/\dhat)}{\Gamma(-\lhat+2i\omega r_H/\dhat)}, \quad \lhat \not\in \mathbb{Z},\mathbb{Z}/2 \vspace*{3mm}\\
\dfrac{2ir_H\omega(-1)^{\lhat}}{\dhat}\prod_{k=1}^{\lhat}\left(k^2+\frac{4r_H^2\omega^2}{\dhat^2}\right), \quad \lhat \in \mathbb{Z}
\end{cases}
\ee
with
\be
W_{\ell\dhat}=\frac{\pi^{\frac{\dhat}{2}+1}r_H^{\dhat(2\lhat+1)}\Gamma\left(-\lhat-\frac{1}{2}\right)}{2^{(4+\dhat)\lhat}\Gamma\left(\lhat+\frac{1}{2}\right)\Gamma\left(\frac{\dhat}{2}(2\lhat+1)\right)}
\ee

More generally, to go 
beyond the case of black holes such as (rotating) neutron stars the calculations requires full numerical studies of the perturbative problem, which can readily be incorporated into the formalism.

Further insights into the information contained in the black hole's response function~\eqref{eq:responsesummary} are revealed by considering limiting cases of particular interest. First, for integer $\lhat$, which applies for four spacetime dimensions, and any frequency within our approximations, the real part of the  response~\eqref{eq:responsesummary} vanishes, hence, the Love numbers are zero, and the purely imaginary terms for $\ell=0$ reduce to the known absorption properties of a black hole~\cite{Page:1976df}. Second, in the static limit $\omega\to 0$, the response~\eqref{eq:responsesummary} reduces to the Love numbers for arbitrary dimensions and multipoles considered in~\cite{Kol:2011vg,Hui:2020xxx}.

\smallskip

Considering wave scattering to calculate Love numbers as done here rather than following the standard approach of working within a stationary setting has two major advantages. (i) Scattering involves imposing boundary conditions both at the horizon and at infinity. The importance of including both of these boundary conditions was shown in~\cite{Chia:2020yla}. Specifically, this bypasses the gauge ambiguities discussed in other studies (e.g. discussed in~\cite{Gralla:2017djj}), which solely consider the near-horizon solution and identify the Love number in terms of the ratio between the growing and decaying solutions for the metric. (ii) By contrast, the waves extracted at null infinity are described by gauge-invariant complex amplitudes, which provide the most convenient identification between the wave solutions of the compact-object perturbation calculations and the skeletonized effective description. In particular, formulating the results in terms of double null coordinates, which have an intrinsic geometric meaning, leads to a clear identification between the two descriptions. One important point to note is that the scalar case we worked out in detail avoids some additional subtleties that we expect to arise in the gravitational case. For instance, we expect the identification of null infinities to only be fixed up to the remaining freedom of supertranslations characterized by the BMS symmetry group~\cite{Bondi:1962px}.

  
\smallskip

The scattering calculations also lead to deeper insights into the necessity and utility of analytic continuations in the multipolar order $\ell$ and the dimension $\dhat$ for different stages of the calculations. Notably, analytic continuations
\begin{enumerate}
 \item immediately distinguish finite size effects from post-Newtonian point-mass terms in quantities characterizing a binary system. Thus, tidal contributions can be unambiguously identified without requiring simultaneous knowledge of the high-order PN point-mass terms having the same scaling with frequency.
\item  have no impact on determining the response function in terms of the wave amplitudes. This is an advantage of considering scattering rather than stationary perturbations.
\item in generic dimension $\dhat$ greatly simplify the calculations. For instance, in the definition of the tortoise coordinate $r^\ast(r)$, a logarithm is present for $\dhat=1$, i.e. $3+1$-dimensional spacetime, which introduces computational subtleties. As discussed in Sec.~\ref{sec:Schwarzschild}, in generic dimensions this relationship is simple, e.g. asymptotically for $r_H\to 0$ it becomes $(r- r^\ast) \sim (r_H/r^\ast)^{\hat{d}} r^\ast +\ldots$. For four spacetime dimensions, the limit is $(r- r^\ast)\sim - r_H\log(r^\ast)+\ldots$.
Thus, it is convenient to work with the simpler general case and obtain the  special cases at the end, similar to the utility of analytic continuations in dimensions in other contexts. 
   \item in generic multipolar order $\ell$ are ubiquitous in analytical black hole perturbation calculations for linking the near-horizon behavior to the asymptotic quantities, e.g.~\cite{Mano:1996vt}. The reason is that for integer values of $\ell$ the two independent near-horizon solutions are degenerate and the single solution is regular, thereby preventing the  identification of the imprints due to the black hole's response from the irregular solution. Analytic continuation to complex angular momentum numbers $\ell$ has also been of great use in related contexts ~\cite{OuldElHadj:2019kji,Folacci:2018sef,Folacci:2019vtt,Folacci:2019cmc}. 
\end{enumerate}

\section{Conclusions and Outlook}
\label{sec:conclusion}

In this paper, we addressed several subtleties and concerns about tidal Love numbers of compact objects. We first considered the problem of identifying the Love numbers in a binary system. We showed that using the gauge-invariant binding energy as a function of frequency for circular orbits, and working in arbitrary dimensions and/or multipolar order, it is straightforward to disentangle high PN order point-particle contributions from finite size effects. We also made explicit the connection between this gauge-invariant energy and tidal coupling coefficients in an effective action. 

Next, we calculated the tidal coupling coefficients and the information about the detailed properties of perturbed compact objects they contain using scattering. This has several advantages over considering stationary perturbations, such as working with quantities defined at null infinity, taking into account all boundary conditions, and gaining insights into the need for and convenience of analytic continuations for different stages of the calculations. We demonstrated the methodology in detail by performing the calculations for scalar perturbations to a Schwarzschild black hole, without specializing to the low-frequency limit as done in most previous works. We  showed that our method recovers known results for the tidal Love numbers and absorption of a black hole in limiting cases. 

Our results represent the basis for a number of future directions. For instance, an important next step is to consider gravitational perturbations. A major simplification arising in the scalar case was that the spacetime remained the background black hole spacetime throughout, which will no longer be true in the gravitational case. Another avenue for future work is to compute the response numerically.
This would enable going beyond the cases where analytical asymptotic expansion are available, for instance generic rotating compact objects.
Our work will be important for future high-precision studies of neutron stars and black holes with gravitational waves, and interpreting the information on fundamental physics encoded in the signals. Further, the methodology established in this paper will also be useful for computing response functions for exotic compact objects and compact objects in alternative theories of gravity, which will yield important information for tests of gravity and beyond-standard-model physics.

\section{Acknowledgments}
G.C. and T.H. acknowledge funding from the Nederlandse Organisatie voor Wetenschappelijk Onderzoek (NWO) sectorplan. 

\appendix
\section{Appendix}\label{sec:appendix}
\subsubsection{Contraction of STF unit vectors in $\dhat$ dimensions}\label{app:ContractionNL}
In this appendix we derive the relation~\eqref{eq:ContrationNL}, i.e. the contraction of two STF unit vectors in any number of dimensions $\dhat\geq1$.
From~\cite{trautman1965lectures} we derive the following result for $\dhat$ dimensions
\begin{align}
	n^{L}=\sum_{k=0}^{[\ell/2]}\frac{(2\ell-2k+\dhat-2)!!}{(2\ell+\dhat-2)!!}\left[\delta^{2k}\tilde{n}^{L-2k}+\text{sym}\right]~.
\end{align}
where ``\text{sym}'' stands for the remaining symmetric terms and we denote non-STF unit vectors by a tilde $\tilde{n}^L$. Upon contracting with a different STF unit vector and using that $n'_{L}n^{L}=\tilde{n}'_{L}n^{L}$, 
\begin{align}
&n'_{L}n^{L}=\tilde{n}'_{L}n^{L}\nonumber\\&=\sum_{k=0}^{[\ell/2]}\frac{(2\ell-2k+\dhat-2)!!}{(2\ell+\dhat-2)!!}\left[\delta^{2k}\tilde{n}^{L-2k}+\text{sym}\right]\tilde{n}'_L\nonumber\\&=\sum_{k=0}^{[\ell/2]}\frac{\ell!}{(\ell-2k)!(2k)!!}\frac{(2\ell-2k+\dhat-2)!!}{(2\ell+\dhat-2)!!}\mu_{nn'}^{L-2k}~.
\end{align}
where we define $\mu_{nn'}\equiv{\tilde{n}\cdot{\tilde{n}'}}$ and in the last equality we contracted all the Kronecker deltas. Next, using the series representation of the $D$ dimensional Legendre Polynomial,
\begin{align}
P_\ell^{(D)}(\mu_{nn'})=\sum_{k=0}^{\infty}(-1)^k\frac{(2\ell-2k+\dhat-2)!!}{(\ell-2k)!(2k)!!(\dhat-2)!!}\mu_{nn'}^{L-2k}
\end{align}
we obtain
\begin{align}
&n'_{L}n^{L}=\frac{\ell!}{(2\ell+\dhat-2)!!}\nonumber\\&\times\sum_{k=0}^{[\ell/2]}\frac{(2\ell-2k+\dhat-2)!!}{(\ell-2k)!(2k)!!}\left(\frac{(\dhat-2)!!}{(\dhat-2)!!}\right)\mu_{nn'}^{L-2k}\nonumber\\&=\frac{\ell!(\dhat-2)!!}{(2\ell+\dhat-2)!!}P_\ell^{(\dhat)}(\mu_{nn'})~.
\end{align}
If $\tilde{n}'=\tilde{n}$, we have $\mu_{nn'}=1$ and therefore
\begin{align}
&n_{L}n^{L}=\frac{\ell!(\dhat-2)!!}{(2\ell+\dhat-2)!!}P_\ell^{(\dhat)}(1)\nonumber\\&=\frac{(\ell+\dhat-1)!(\dhat-2)!!}{(2\ell+\dhat-2)!!(\dhat-1)!}\nonumber\\&=\frac{(\ell+\dhat-1)!}{(2\ell+\dhat-2)!!(\dhat-1)!!}~.
\end{align}
where we have used that $(\dhat-1)!=(\dhat-1)!!(\dhat-2)!!$ and \cite{CunhaCampos:2012}
\begin{align}
P_\ell^{(\dhat)}(1)=\frac{(\ell+\dhat-1)!}{\ell!(\dhat-1)!}~.
\end{align}
Remarkably, this result is valid for $\dhat\geq1$ and reduces to the known four-dimensional values for $\dhat=1$.

\subsubsection{Distributional Laplacian}\label{app: DistrLaplacian}
\label{ssec:DistrLaplacian}
From~\cite{Generalizedfunctions}, the (spatial)  D dimensional distributional partial derivative of any homogenous function $f(x)$ of degree $\lambda$, i.e. a function such that $f(ax)=a^\lambda{f(x)}$ for $a>0$, is given by
\begin{align}
\tilde{\p}_if(x)=&\p_if(x)+\frac{(-1)^k}{k!}\frac{\tilde{\p}^k}{\tilde{\p}{x^{\ell_1}}\tilde{\p}{x^{\ell_2}}\dots\tilde{\p}{x^{\ell_k}}}\delta(x)\nonumber\\&\times\oint_{\mathcal{S}^{D-1}}d\mathcal{S}n^i f(x') x'^{\ell_1}\dots{x^{\ell_k}}
\end{align}
where the derivative with a tilde means a distributional derivative and $k=-\lambda-D+1>0$. Since we will use this formula to compute derivatives of inverse powers, we will make the definition $\alpha\equiv-\lambda$ such that $k=\alpha-D+1$. In particular we want to compute the distributional Laplacian of $r^{-\alpha}$, so we begin by computing the first derivative,
\begin{align}
\tilde{\p}_i\left(\frac{1}{r^\alpha}\right)=&{\p}_i\left(\frac{1}{r^\alpha}\right)+\frac{(-1)^{\alpha-D+1}}{(\alpha-D+1)!}\tilde{\p}_{\alpha-D+1}\delta(x)\nonumber\\&\times\oint_{\mathcal{S}^{D-1}}d\mathcal{S}n'^i \left(\frac{1}{r'^\alpha}\right)x'^{\ell_1}\dots x'^{\ell_{\alpha-D+1}}~.
\end{align}
The closed surface integral is given by
\begin{align}
&\oint_{\mathcal{S}^{D-1}}d\mathcal{S'}n'_i \left(\frac{1}{r'^\alpha}\right)x'^{i_1}\dots x'^{\alpha-D+1}\nonumber\\&=\oint_{\mathcal{S}^{D-1}}d\Omega_{D-1}r'^{D-1}n'^in'^{\ell_1}\dots n'^{\ell_{\alpha-D+1}}\left(\frac{1}{r'^{D-1}}\right)\nonumber\\&=\oint_{\mathcal{S}^{D-1}}d\Omega_{D-1}n'^in'^{\ell_1}\dots n'^{\ell_{\alpha-D+1}}\nonumber\\&=\frac{(D-2)!!}{\alpha!!}\Omega_{D-1}\delta_{\{i\ell_1\dots}\delta_{\ell_{\alpha-D}\ell_{\alpha-D+1}\}}\delta_{\alpha-D,2n}
\end{align}
where we introduced the Kronecker delta $\delta_{\alpha-D,2n}$ with $n$ an integer to account for the fact that $\alpha-D$ has to be even and we used the well-known property
\begin{align}
&\oint_{\mathcal{S}^{D-1}}d\mathcal{S'}n'^{i_1}\dots x'^{i_{2m}}\nonumber\\&=\frac{(D-2)!!}{(D+2m-2)!!}\Omega_{D-1}\delta_{\{i\ell_1\dots}\delta_{\ell_{2m-1}\ell_{2m}\}}~,
\end{align}
with $\Omega_{D-1}=2\pi^{D/2}/\Gamma(D/2)$. The first distributional derivative is, taking into account that $k>0$,
\begin{align}
\tilde{\p}_i&\left(\frac{1}{r^\alpha}\right)={\p}_i\left(\frac{1}{r^\alpha}\right)\nonumber\\&+\frac{(-1)^{\alpha-D+1}}{(\alpha-D+1)!}\frac{(D-2)!!}{\alpha!!}\Omega_{D-1}\delta_{\{i\ell_1\dots}\delta_{\ell_{\alpha-D}\ell_{\alpha-D+1}\}}\nonumber\\&\times\delta_{\alpha-D,2n}\Theta(\alpha-D+1)\tilde{\p}_{\alpha-D+1}\delta(x)~.
\end{align}
Now we can compute the second derivative,
\begin{align}\label{eq:doubledistrderiv}
\tilde{\p}_j&\tilde{\p}_i\left(\frac{1}{r^\alpha}\right)=\tilde{\p}_j{\p}_i\left(\frac{1}{r^\alpha}\right)\nonumber\\&+\frac{(-1)^{\alpha-D+1}}{(\alpha-D+1)!}\frac{(D-2)!!}{\alpha!!}\Omega_{D-1}\delta_{\{i\ell_1\dots}\delta_{\ell_{\alpha-D}\ell_{\alpha-D+1}\}}\nonumber\\&\times\delta_{\alpha-D,2n}\Theta(\alpha-D+1)\tilde{\p}_j\tilde{\p}_{ \alpha-D+1}\delta(x),
\end{align}
with
\begin{align}
\tilde{\p}_j&{\p}_i\left(\frac{1}{r^\alpha}\right)={\p}_j{\p}_i\left(\frac{1}{r^\alpha}\right)\nonumber\\&-\alpha\frac{(-1)^{\alpha-D+2}}{(\alpha-D+2)!}\tilde{\p}_{\alpha-D+2}\delta(x)\nonumber\\&\times\oint_{\mathcal{S}^{D-1}}d\mathcal{S}n'^j \left(\frac{n^i}{r'^{\alpha+1}}\right)x'^{\ell_1}\dots x'^{\ell_{\alpha-D+2}}~.
\end{align}
where we used that $k=\alpha+1-D+1=\alpha-D+2$ since
\begin{align}
\p_i\left(\frac{1}{r^\alpha}\right)=-\alpha\frac{1}{r^{\alpha+1}}n_i~.
\end{align}
The angular integral is given by
\begin{align}
&\oint_{\mathcal{S}^{D-1}}d\mathcal{S}n'^j \left(\frac{n^i}{r'^{\alpha+1}}\right)x'^{\ell_1}\dots x'^{\ell_{\alpha-D+2}}\nonumber\\&=\oint_{\mathcal{S}^{D-1}}d\Omega_{D-1}n'^jn'^in'^{\ell_1}\dots n'^{\ell_{\alpha-D+2}}\nonumber\\&=\frac{(D-2)!!}{(\alpha+2)!!}\Omega_{D-1}\delta_{\{ij\dots}\delta_{\ell_{\alpha-D+1}\ell_{\alpha-D+2}\}}\delta_{\alpha-D,2n}~,
\end{align}
such that
\begin{align}
\tilde{\p}_j&\tilde{\p}_i\left(\frac{1}{r^\alpha}\right)={\p}_j{\p}_i\left(\frac{1}{r^\alpha}\right)\nonumber\\&+\frac{(-1)^{\alpha-D+1}}{(\alpha-D+1)!}\frac{(D-2)!!}{\alpha!!}\Omega_{D-1}\delta_{\{i\ell_1\dots}\delta_{\ell_{\alpha-D}\ell_{\alpha-D+1}\}}\nonumber\\&\times\delta_{\alpha-D,2n}\Theta(\alpha-D+1)\tilde{\p}_j\tilde{\p}_{ \alpha-D+1}\delta(x)\nonumber\\&-\alpha\frac{(-1)^{\alpha-D+2}}{(\alpha-D+2)!}\frac{(D-2)!!}{(\alpha+2)!!}\Omega_{D-1}\delta_{\{ij\dots}\delta_{\ell_{\alpha-D+1}\ell_{\alpha-D+2}\}}\nonumber\\&\times\Theta(\alpha-D+2)\delta_{\alpha-D,2n}\tilde{\p}_{\alpha-D+2}\delta(x)~\label{eq:seconddistrderivative}.
\end{align}
Now, in order to obtain the Laplacian we just have to contract the indices, or equivalently introduce a Kronecker delta. When doing that, we will have to contract the Kronecker delta with other symmetrized Kronecker deltas that will give the same contribution. Specifically,
\begin{align}
\delta^{ij}&\delta_{\{ij\dots}\delta_{\ell_{\alpha-D+1}\ell_{\alpha-D+2}\}}=\delta^{ij}\delta_{ij}\delta_{\{\ell_1\ell_2}\dots\delta_{\ell_{\alpha-D+1}\ell_{\alpha-D+2}\}}\nonumber\\&+\delta^{ij}\delta_{\{i\ell_1}\delta_{j\ell_2}\dots\delta_{\ell_{\alpha-D+1}\ell_{\alpha-D+2}\}}{}_{/ij}\nonumber\\&=D(\alpha-D+1)\delta_{\ell_1\ell_2}\dots\delta_{\ell_{\alpha-D+1}\ell_{\alpha-D+2}}\nonumber\\&+\delta^{ij}2(\alpha-D+1)!\delta_{i\ell_1}\delta_{j\ell_2}\dots\delta_{\ell_{\alpha-D+1}\ell_{\alpha-D+2}}\label{eq:deltaijsymm}
\end{align}
where $/ij$ means all combinations but the one with $ij$. The factors in front of the unsymmetrized deltas come from taking as many combinations of 2 of free indices and dividing by the number of Kronecker deltas one can form with those indices (for instance the first term of~\eqref{eq:deltaijsymm}, is $(\alpha-D+2)!/(2!(\alpha-D)!)\times2/(\alpha-D+2)$). So far we have treated the case $\alpha-D+2>0$. In order to compute the limiting case one should go back to~\eqref{eq:doubledistrderiv} and just keep the first term with $\alpha=D-2$. Finally, the distributional Laplacian of $r^{-\alpha}$ for generic dimensions is given by 
\begin{widetext}
\label{eq:distrLaplaciancases}
\begin{align}
\tilde{\nabla}^2\left(\frac{1}{r^\alpha}\right)=\begin{cases}{\nabla}^2\left(\dfrac{1}{r^\alpha}\right)+\dfrac{(-1)^{\alpha+D-3}}{(\alpha-D+2)!}\dfrac{(D-2)!!}{(\alpha+2)!!}\delta_{\alpha-D,2n}\Omega_{D-1}\tilde{\nabla}^2\delta(x)\vspace*{1mm}\\\quad\times\Big[2(\alpha-D+1)!\left(\alpha\Theta(\alpha-D+2)+(\alpha+2)\Theta(\alpha-D+1)\right)+\alpha{D}(\alpha-D+1)\Theta(\alpha-D+2)\Big]~,\,\,\alpha>D-2\vspace*{4mm}\\
{\nabla}^2\left(\dfrac{1}{r^\alpha}\right)-(D-2)\Omega_{D-1}\delta(x)~,\,\,\alpha=D-2\vspace*{4mm}\\
{\nabla}^2\left(\dfrac{1}{r^\alpha}\right)~,\,\,\alpha<D-2
\end{cases}
\end{align}
Recall that in the main text we use $\dhat=D-2$ instead of $D$.
\end{widetext}

\subsubsection{Extracting the finite part of the tidal term}\label{app:AppendixFP}
In order to compute the response function we have to extract the finite part of the tidal term, $\p_L\phi$. For that, we will directly substitute the series representation of the Bessel functions and apply the STF derivatives and their identities. From \eqref{eq:varphi} we obtain
\begin{align}
\p_L\phi
=\sum_{k=0}^\infty\left({C}_{\rm reg}^K\p_L\p_K\phi^{(0)}_{\rm reg}+{C}_{\rm irreg}^K\p_L\p_K\phi^{(0)}_{\rm irreg}\right),
\end{align}
where
\begin{subequations}
\begin{align}
\p_L\p_K\phi^{(0)}_{\rm reg}&=e^{i\omega{t}}\sqrt{2\pi\omega}~\p_L\p_K\left(r^{-\dhat/2}J_{\dhat/2}(\omega r)\right)~,\\
\p_L\p_K\phi^{(0)}_{\rm irreg}&=e^{i\omega{t}}\sqrt{2\pi\omega}~\p_L\p_K\left(r^{-\dhat/2}Y_{\dhat/2}(\omega r)\right)~.
\end{align} 
\end{subequations}
We begin with the regular piece,
\begin{align}
&\p_L\p_K\left(r^{-\dhat/2}J_{\dhat/2}(\omega r)\right)\\
&=\sum_{m=0}^\infty\frac{(-1)^m}{m!\Gamma(m+\frac{\dhat}{2}+1)}\left(\frac{\omega}{2}\right)^{2m+\dhat/2}\p_L\p_K\left(r^{2m}\right)~.
\end{align}
In order to obtain the finite part we have to take $2m=\ell+k$ derivatives. Using (A14) of \cite{1994CeMDA..60..139H},
\be\label{eq:STFidentity1}
\p_P\left(r^{2j}\right)=0\qquad\text{if}~ j=0,1,2,\dots,p-1
\ee
implies that in order to have a non-zero result  $2m=\ell+k\geq\ell$ and $2m=\ell+k\geq{k}$. Therefore, the only possible choice is $\ell=k$ for which $m=\ell$. Using (A13) and (A12) of \cite{1994CeMDA..60..139H},
\begin{align}
&\p_P\left(r^{\kappa}\right)=\frac{\kappa!!}{(\kappa-2p)!!}n_P~r^{\kappa-p}~,\label{eq:STFidentity2}\\
&\p_iX_P=p~\delta_{i<i_p} X_{P-1>}~\label{eq:STFidentity3},
\end{align}
yields
\be\label{eq:finitepartr}
\p_L\p_L\left(r^{2\ell}\right)=(2\ell)!!\p_L\left(n_L r^{\ell}\right)=\ell!(2\ell)!!~.
\ee 
Hence,
\begin{align}\label{eq:finitepartJplus}
\underset{r\rightarrow{0}}{\text{FP}}\p_L\p_K\left(r^{-\dhat/2}J_{\dhat/2}(\omega r)\right)=\frac{\ell!~2^\ell(-1)^\ell}{\Gamma(\frac{\dhat}{2}+\ell+1)}\left(\frac{\omega}{2}\right)^{2\ell+\dhat/2}~,
\end{align}
where we have used that $(2\ell)!!=2^\ell\ell!$. Similarly, we can compute the finite part of the irregular solution. Recall that the Bessel function of the second kind reads 
\begin{align}\label{eq:BesselfunctionYapp}
Y_{\dhat/2}(\omega{r})=\frac{1}{\sin\left(\frac{\pi\dhat}{2}\right)}\left[\cos\left(\frac{\pi\dhat}{2}\right)J_{\dhat/2}(\omega{r})-J_{-\dhat/2}(\omega{r})\right]~.
\end{align}
The first term is proportional to the regular solution and therefore we will focus on the second term,
\begin{align}\label{eq:partialJminus}
&\p_L\p_K\left(r^{-\dhat/2}J_{-\dhat/2}(\omega r)\right)\\
&=\sum_{m=0}^\infty\frac{(-1)^m}{m!\Gamma(m-\frac{\dhat}{2}+1)}\left(\frac{\omega}{2}\right)^{2m-\dhat/2}\p_L\p_K\left(r^{2m-\dhat}\right)~.
\end{align}
Now the condition to have a non-zero result reads $2m-\dhat=\ell+k$. Using \eqref{eq:STFidentity1} implies $2m-\dhat=\ell+k\geq\ell$ and $2m-\dhat=\ell+k\geq{k}$ and therefore $m=\dhat/2+\ell$. Plugging \eqref{eq:finitepartr} back into \eqref{eq:partialJminus} yields

\begin{align}\label{eq:finitepartJminus}
\underset{r\rightarrow{0}}{\text{FP}}&\p_L\p_K\left(r^{-\dhat/2}J_{-\dhat/2}(\omega r)\right)\\&=\frac{\ell!~2^\ell(-1)^\ell\cos\left(\frac{\pi\dhat}{2}\right)}{\Gamma(\frac{\dhat}{2}+\ell+1)}\left(\frac{\omega}{2}\right)^{2\ell+\dhat/2}~,
\end{align}
where given that $m=\dhat/2+\ell$ is an integer and $\dhat$ can be odd or even,
\begin{align}
(-1)^{\dhat/2}=\cos\left(\frac{\pi\dhat}{2}\right)~.
\end{align}
Combining \eqref{eq:finitepartJplus} and \eqref{eq:finitepartJminus} into \eqref{eq:BesselfunctionYapp} yields
\begin{align}
\underset{r\rightarrow{0}}{\text{FP}}\p_L\p_K\phi_{\rm irreg}^{(0)}\propto\underset{r\rightarrow{0}}{\text{FP}}\p_L\p_K\left(r^{-\dhat/2}Y_{\dhat/2}(\omega r)\right)=0~.
\end{align}
We can now compute the frequency-dependent tidal field
\begin{align}
E_L(\omega)&=\underset{r\rightarrow{0}}{\text{FP}}\p_L\phi(\omega)={C}_{\rm reg}^L\underset{r\rightarrow{0}}{\text{FP}}\p_L\p_L\phi_{\rm reg}^{(0)}(\omega)\nonumber\\=&e^{i\omega{t}}\ell!\pi\left(\frac{\omega}{2}\right)^{\dhat/2+1/2+2\ell}\frac{(-1)^{\ell}2^{\ell+1}}{\Gamma(\frac{\dhat}{2}+\ell+1)}{C}_{\rm reg}^L~.
\end{align}
where we use that $\phi(\omega)=\sqrt{2\pi}e^{-i\omega t}\phi(t)$ for a fixed frequency $\omega$.

\bibliography{refs_ScatterEFT}
\end{document}